\documentclass{lmcs}
\pdfoutput=1

\usepackage{lastpage}
\lmcsdoi{21}{1}{6}
\lmcsheading{}{\pageref{LastPage}}{}{}%
{Nov.~28,~2023}{Jan.~22,~2025}{}

\usepackage[utf8]{inputenc}
\usepackage{amsmath,
	    amsfonts,
	    amstext,
	    amssymb,
	    mathrsfs,
	    stmaryrd,
	    latexsym,
	    enumitem,
	    proof,
	    graphicx,
	    color,
	    hyperref,
	    comment,
	    semantic, 
	    tikz}

\usetikzlibrary{arrows,positioning}
\tikzset{
modal/.style={
>=stealth',shorten >=0pt,shorten <=0pt,auto,
node distance=1cm,semithick}, 
point/.style={circle,draw,fill=black,inner sep=0.5mm},
}

\newcommand{\ms}[1]
	{\null\ifmmode\mathord{\mathcode`-="702D\it #1\mathcode`\-="2200}
	\else$\mathord{\mathcode`-="702D\it #1\mathcode`\-="2200}$\fi}

\newcommand{\cws}[2]
	{\\ \centerline{$#2$} \\[-#1pt]}

\newcommand{\fullbox}
	{{\mbox{}\nolinebreak\hfill{$\rule{2mm}{2mm}$}}}

\newcommand{\bibtrick}[1]
	{}

\newcommand{\cala}
        {\mathcal{A}}

\newcommand{\calb}
        {\mathcal{B}}

\newcommand{\calh}
        {\mathcal{H}}

\newcommand{\call}
        {\mathcal{L}}

\newcommand{\calp}
        {\mathcal{P}}

\newcommand{\calq}
        {\mathcal{Q}}

\newcommand{\calr}
        {\mathcal{R}}

\newcommand{\cals}
        {\mathcal{S}}

\newcommand{\natns}
	{\mathbb{N}}

\newcommand{\procs}
	{\mathbb{P}}

\newcommand{\first}
	{\textit{first}}

\newcommand{\last}
	{\textit{last}}

\newcommand{\pt}
	{\textit{path}}

\newcommand{\Path}
	{\textit{Path}}

\newcommand{\Run}
	{\textit{Run}}

\newcommand{\arrow}[2]
        {\, {\auxarrow\limits^{#1}}_{#2} \,}
\newcommand{\auxarrow}
	{\mathop{\longrightarrow}}

\newcommand{\warrow}[2]
        {\, {\wauxarrow\limits^{#1}}_{#2} \,}
\newcommand{\wauxarrow}
	{\mathop{= \!\!\!\! \Longrightarrow}}

\newcommand{\nil}
	{\underline 0}

\newcommand{\eqdef}
	{\triangleq}

\newcommand{\sbis}[1]
	{\sim_{#1}}

\newcommand{\wbis}[1]
        {\approx_{#1}}

\newcommand{\pco}[1]
	{\mathop{\Vert_{#1}}}

\newcommand{\reach}
	{\textit{reach}}

\begin{document}

\title[Noninterference Analysis of Reversible Systems]
      {Noninterference Analysis of Reversible Systems: \texorpdfstring{\\}{}
       {An Approach Based on Branching Bisimilarity}}

\author[A.~Esposito]{Andrea Esposito\lmcsorcid{0009-0009-2259-902X}}[a]
\author[A.~Aldini]{Alessandro Aldini\lmcsorcid{0000-0002-7250-5011}}[a]
\author[M.~Bernardo]{\texorpdfstring{\\}{}Marco Bernardo\lmcsorcid{0000-0003-0267-6170}}[a]
\address{Dipartimento di Scienze Pure e Applicate, Universit\`a di Urbino, Italy}
\email{andrea.esposito@uniurb.it, alessandro.aldini@uniurb.it, marco.bernardo@uniurb.it}

\author[S.~Rossi]{Sabina Rossi\lmcsorcid{0000-0002-1189-4439}}[b]
\address{Dipartimento di Scienze Ambientali, Informatica e Statistica, Universit\`a Ca' Foscari, Venezia,
	 Italy}
\email{sabina.rossi@unive.it}



\begin{abstract}
The theory of noninterference supports the analysis of information leakage and the execution of secure
computations in multi-level security systems. Classical equivalence-based approaches to noninterference
mainly rely on weak bisimulation semantics. We show that this approach is not sufficient to identify
potential covert channels in the presence of reversible computations. As illustrated via a database
management system example, the activation of backward computations may trigger information flows that are
not observable when proceeding in the standard forward direction. To capture the effects of back-and-forth
computations, it is necessary to switch to a more expressive semantics, which has been proven to be
branching bisimilarity in a previous work by De Nicola, Montanari, and Vaandrager. In this paper we
investigate a taxonomy of noninterference properties based on branching bisimilarity along with their
preservation and compositionality features, then we compare it with the taxonomy of Focardi and Gorrieri
based on weak bisimilarity.
\end{abstract}

\maketitle

%
%
\section{Introduction}
\label{sec:intro}
%
%

Noninterference was introduced by Goguen and Meseguer~\cite{GM82} to reason about the way in which
illegitimate information flows can occur due to covert channels from high-level agents to low-level ones in
multi-level security systems. Since the first definition, conceived for deterministic state machines, in the
last four decades a lot of work has been done that led to a variety of extensions (dealing with
nondeterminism or quantitative domains) in multiple frameworks (from language-based security to concurrency
theory); see, e.g., \cite{FG01,Ald06,Man11,HS12,ABG04,HMPR21} and the references therein. Analogously, the
techniques proposed to verify information-flow security properties based on noninterference have followed
several different approaches, ranging from the application of type theory~\cite{ZM04} and abstract
interpretation~\cite{GM18} to control flow analysis and equivalence or model
checking~\cite{FPR02,Mar03,AB11}.

Noninterference guarantees that low-level agents cannot infer from their observations what high-level ones
are doing. Regardless of its specific definition, noninterference is closely tied to the notion of
behavioral equivalence~\cite{Gla01}, because the idea is to compare the system behavior with high-level
actions being prevented and the system behavior with the same actions being hidden. One of the most
established formal definitions of noninterference properties relies on weak bisimilarity in a process
algebraic framework~\cite{Mil89a}, as it naturally lends itself to reason formally about covert channels and
illegitimate information~flows.

After the classification of security properties in~\cite{FG01}, the literature concentrated on weak
bisimilarity. Here we claim that it is worth studying nondeterministic noninterference in a different
setting, relying on branching bisimulation semantics. This was introduced in~\cite{GW96} as a refinement of
weak bisimilarity to preserve the branching structure of processes also when abstracting from unobservable
actions. It features a complete axiomatization whose only $\tau$-axiom is $a \, . \, (\tau \, . \, (x + y) +
x) = a \, . \, (x + y)$, where dot stands for action prefix, plus stands for nondeterministic choice, $a$ is
an action, $\tau$ is an unobservable action, and $x$ and $y$ \linebreak are process terms. Moreover, while
weak bisimilarity can be verified in $O(n^2 \cdot m \cdot \log n)$, where $m$ is the number of transitions
and $n$ is the number of states of the labeled transition system underlying the process at hand, branching
bisimilarity can be verified more efficiently. An $O(m \cdot n)$ algorithm was provided in~\cite{GV90} and,
more recently, an even faster $O(m \cdot \log n)$ algorithm has been developed in~\cite{JGKW20}. 

A clear motivation for switching from weak to branching bisimilarity is provided by reversible processes,
whose computational model features both forward and backward computations~\cite{Lan61,Ben73}. Reversible
computing has turned out to have interesting applications in biochemical reaction
modeling~\cite{PUY12,Pin17}, parallel discrete-event simulation~\cite{PP14,SOJB18}, robotics~\cite{LES18},
control theory~\cite{SPP19}, wireless communications~\cite{SPP19}, fault-tolerant
systems~\cite{DK05,VKH10,LLMSS13,VS18}, and concurrent program debugging~\cite{GLM14,LNPV18a}. To the best
of our knowledge, no information flow security approach exists for this setting, in which we will see that
weak bisimilarity does not represent a proper tool for the comprehensive analysis of covert channels.
 
Behavioral equivalences for reversible processes must take into account the fact that computations are
allowed to proceed not only forward, but also backward. To this aim, back-and-forth bisimilarity, introduced
in~\cite{DMV90}, requires that two systems are able to mimic each other's behavior stepwise not only in
performing actions that follow the arrows of the labeled transition systems, but also in undoing those
actions when going backward. Formally, back-and-forth bisimulations are defined on computation paths instead
of states thus preserving not only causality but also history, as backward moves are constrained to take
place along the same path followed in the forward direction even in the presence of concurrency.
In~\cite{DMV90} it was shown that strong back-and-forth bisimilarity coincides with the usual notion of
strong bisimilarity, while weak back-and-forth bisimilarity is surprisingly finer than standard weak
bisimilarity, and it coincides with branching bisimilarity. In particular, this latter result will allow us
to investigate the nature of covert channels in reversible systems by using a standard process calculus,
i.e., without having to decorate executed actions like in~\cite{PU07,BR23} or store them into stack-based
memories like in~\cite{DK04}.

Once established that branching bisimilarity enables noninterference analysis of reversible systems, the
novel contribution of this paper is the study of noninterference security properties based on branching
bisimilarity. In addition to investigating preservation and compositionality features, we compare the
resulting properties with those based on weak bisimilarity~\cite{FG01} and we develop a taxonomy of the
former that can be naturally applied to those based on weak back-and-forth bisimilarity for reversible
systems. Moreover, we illustrate how, in the setting of reversible systems, weak bisimilarity does not
provide a proper framework for the identification of subtle covert channels, while branching bisimilarity
does. This is carried out through a database management system example. 

This paper, which is a revised and extended version of~\cite{EAB23}, is organized as follows. In
Section~\ref{sec:basic_defs} we recall background definitions and results for several bisimulation
equivalences as well as a number of information-flow security properties based on weak bisimilarity, along
with a process language to formalize those properties. In Section~\ref{sec:example_p1} we introduce the
database management system example. In Section~\ref{sec:branching_properties}, after recasting the same
information-flow security properties in terms of branching bisimilarity, we present some results about the
preservation of those properties under branching bisimilarity and their compositionality with respect to the
operators of the considered language, then we study the relationships among all the previously discussed
properties and summarize them in a new taxonomy. \linebreak In Section~\ref{sec:bf} we recall the notion of
back-and-forth bisimulation and its relationship with the aforementioned bisimulations, emphasizing that
weak back-and-forth bisimilarity coincides with branching bisimilarity, which allows us to apply our results
to reversible systems. \linebreak In Section~\ref{sec:example_p2} we add reversibility to the database
management system example to illustrate the need of branching-bisimilarity-based noninterference. Finally,
in Section~\ref{sec:concl} we provide some concluding remarks and directions for future work.

%
%
\section{Background Definitions and Results}
\label{sec:basic_defs}
%
%

In this section we recall bisimulation equivalences (Section~\ref{sec:bisim}) and introduce a basic process
language (Section~\ref{sec:proc_lang}) through which we will express bisimulation-based information-flow
security properties (Section~\ref{sec:weak_bisim_properties}).

%
\subsection{Bisimulation Equivalences}
\label{sec:bisim}
%

Bisimilarity is one of the most important behavioral equivalences~\cite{Gla01}. To represent the behavior of
a process, we use a labeled transition system~\cite{Kel76}, which is a state-transition graph whose
transitions are labeled with actions.

	\begin{defi}\label{def:lts}

A \emph{labeled transition system (LTS)} is a triple $(\cals, \cala_{\tau}, \! \arrow{}{} \!)$ where $\cals$
is a nonempty, at most countable set of states, $\cala_{\tau} = \cala \cup \{ \tau \}$ is a countable set of
actions including an unobservable action denoted by $\tau$, and $\! \arrow{}{} \! \subseteq \cals \times
\cala_{\tau} \times \cals$ is a transition relation. \fullbox

	\end{defi}
\noindent 
A transition $(s, a, s')$ is written $s \arrow{a}{} s'$, where $s$ is the source state, $a$ is the
transition label, and $s'$ is the target state, in which case we say that $s'$ is reachable from $s$ via
that $a$-transition. In general, we say that $s'$ is reachable from $s$, written $s' \in \reach(s)$, iff $s'
= s$ or there is a finite sequence of transitions such that the target state of each of them coincides with
the source state of the subsequent one, with the source of the first one being~$s$ and the target of the
last one being $s'$.

Strong bisimilarity~\cite{Par81,Mil89a} identifies processes that are able to mimic each other's behavior
stepwise. Unlike trace equivalence, only processes with the same branching structure can be equated. For
instance, $a \, . \, (x + y)$ and $a \, . \, x + a \, . \, y$ are told apart unless $x = y$.

	\begin{defi}\label{def:bisim}

Let $(\cals, \cala_{\tau}, \! \arrow{}{} \!)$ be an LTS and $s_{1}, s_{2} \in \cals$. We say that $s_{1}$
and $s_{2}$ are \emph{strongly bisimilar}, written $s_{1} \sbis{} s_{2}$, iff $(s_{1}, s_{2}) \in \calb$ for
some strong bisimulation $\calb$. \linebreak A symmetric binary relation $\calb$ over $\cals$ is a
\emph{strong bisimulation} iff, whenever $(s_{1}, s_{2}) \in \calb$, then:

		\begin{itemize}

\item for each $s_{1} \arrow{a}{} s'_{1}$ there exists $s_{2} \arrow{a}{} s'_{2}$ such that $(s'_{1},
s'_{2}) \in \calb$.
\fullbox

		\end{itemize}

	\end{defi}
\noindent 
Weak bisimilarity~\cite{Mil89a} is coarser than strong bisimilarity because it is capable of abstracting
from unobservable actions, which are denoted by~$\tau$. As an example, $a \, . \, \tau \, . \, x = a \, . \,
x$. Let $s \warrow{\tau^{*}}{} s'$ mean that $s' \in \reach(s)$ and, whenever $s' \neq s$, there is a finite
sequence of transitions that starts in $s$ and terminates in $s'$, where each transition is labeled with
$\tau$. Moreover, let $\warrow{\tau^{*}}{} \! \arrow{a}{} \! \warrow{\tau^{*}}{}$ mean that an
$a$-transition is possibly preceded and followed by finitely many $\tau$-transitions.

	\begin{defi}\label{def:weak_bisim}

Let $(\cals, \cala_{\tau}, \! \arrow{}{} \!)$ be an LTS and $s_{1}, s_{2} \in \cals$. We say that $s_{1}$
and $s_{2}$ are \emph{weakly bisimilar}, written $s_{1} \wbis{} s_{2}$, iff $(s_{1}, s_{2}) \in \calb$ for
some weak bisimulation~$\calb$. A symmetric binary relation $\calb$ over $\cals$ is a \emph{weak
bisimulation} iff, whenever $(s_{1}, s_{2}) \in \calb$, then:

		\begin{itemize}

\item for each $s_{1} \arrow{\tau}{} s'_{1}$ there exists $s_{2} \warrow{\tau^{*}}{} s'_{2}$ such that
$(s'_{1}, s'_{2}) \in \calb$;

\item for each $s_{1} \arrow{a}{} s'_{1}$ with $a \in \cala$ there exists $s_{2} \warrow{\tau^{*}}{} \!
\arrow{a}{} \! \warrow{\tau^{*}}{} s'_{2}$ such that $(s'_{1}, s'_{2}) \in \calb$.
\fullbox

		\end{itemize}

	\end{defi}
\noindent 
Branching bisimilarity~\cite{GW96}, which is coarser than strong bisimilarity too, is finer than weak
bisimilarity because it preserves the branching structure of processes even when abstracting from
$\tau$-actions -- see the condition $(s_{1}, \bar{s}_{2}) \in \calb$ in the definition below.

	\begin{defi}\label{def:branching_bisim}

Let $(\cals, \cala_{\tau}, \! \arrow{}{} \!)$ be an LTS and $s_{1}, s_{2} \in \cals$. We say that $s_{1}$
and~$s_{2}$ are \emph{branching bisimilar}, written $s_{1} \wbis{\rm b} s_{2}$, iff $(s_{1}, s_{2}) \in
\calb$ for some branching bisimulation $\calb$. \linebreak A symmetric binary relation $\calb$ over $\cals$
is a \emph{branching bisimulation} iff, whenever $(s_{1}, s_{2}) \in \calb$, then:

		\begin{itemize}

\item for each $s_{1} \arrow{a}{} s'_{1}$:

			\begin{itemize}

\item either $a = \tau$ and $(s'_{1}, s_{2}) \in \calb$;

\item or there exists $s_{2} \warrow{\tau^{*}}{} \bar{s}_{2} \arrow{a}{} s'_{2}$ such that $(s_{1},
\bar{s}_{2}) \in \calb$ and $(s'_{1}, s'_{2}) \in \calb$.
\fullbox

			\end{itemize}

		\end{itemize}

	\end{defi}

	\begin{figure}[t]

\begin{center}
\begin{tikzpicture}[modal]

\node[point] (root) [label = 90: {$s_1$}]                   {};
\node[point] (p1)   [below left = of root, label = 180: {}] {};
\node[point] (p2)   [below= of root, label = 180: {}]       {};
\node[point] (p3)   [below right = of root, label = 0: {}]  {};
\node[point] (p4)   [below= of p1, label = 180: {}]         {};

\path[->] (root) edge node[left]  {$\tau$}              (p1);
\path[->] (root) edge node[right] {$b$}                 (p3);
\path[->] (root) edge node[right] {$\hspace{-0.1cm} a$} (p2);
\path[->] (p1)   edge node[left]  {$a \hspace{-0.1cm}$} (p4);

\node[point] (qroot) [right = 4cm of root, label = 90: {$s_2$}] {};
\node[point] (q1)    [below left = of qroot, label = 180: {}]   {};
\node[point] (q2)    [below right = of qroot, label = 0: {}]    {};
\node[point] (q3)    [below= of q1, label = 180: {}]            {};

\path[->] (qroot) edge node[left]  {$\tau$}              (q1);
\path[->] (qroot) edge node[right] {$b$}                 (q2);
\path[->] (q1)    edge node[left]  {$a \hspace{-0.1cm}$} (q3);

\end{tikzpicture}
\end{center}

\caption{States $s_1$ and $s_2$ are weakly bisimilar but not branching bisimilar}
\label{fig:wb_brb_cex}

	\end{figure}
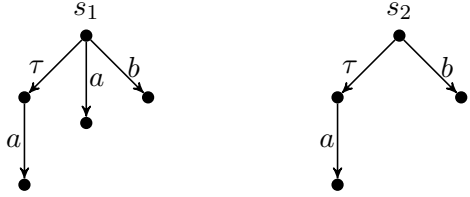
\noindent 
An example that highlights the higher distinguishing power of branching bisimilarity is given in
Figure~\ref{fig:wb_brb_cex}, where the LTS is depicted as a directed graph in which vertices represent
states and action-labeled edges represent transitions. The states $s_1$ and $s_2$ are weakly bisimilar but
not branching bisimilar. The only transition that distinguishes $s_1$ from $s_2$ is the $a$-transition
of~$s_1$, which can be mimicked by $s_2$ according to weak bisimilarity by performing its $\tau$-transition
followed by its $a$-transition. However, $s_2$ cannot respond in the same way according to branching
bisimilarity. If $s_2$ performs the $\tau$-transition followed by the $a$-transition, then the state reached
after the $\tau$-transition should be branching bisimilar to $s_1$, which is not the case because of the
$b$-transition departing from $s_1$.

%
\subsection{A Process Language with High and Low Actions}
\label{sec:proc_lang}
%

We now introduce a basic process calculus to formalize the security properties of interest. To address two
security levels, actions are divided into high and low. We partition the set $\cala$ of observable actions
into $\cala_\calh \cup \cala_\call$, with $\cala_\calh \cap \cala_\call = \emptyset$, where $\cala_\calh$ is
the set of high-level actions, ranged over by~$h$, and $\cala_\call$ is the set of low-level actions, ranged
over by $l$. Note that $\tau \notin \cala_\calh \cup \cala_\call$.

The set $\procs$ of process terms is obtained by considering typical operators from CCS~\cite{Mil89a} and
CSP~\cite{BHR84}. In addition to prefix and choice, we have restriction and hiding as they are necessary to
formalize noninterference properties, the CSP parallel composition so as not to turn into~$\tau$ the
synchronization between high-level actions as would happen with the CCS parallel composition, and recursion
(which was not considered in~\cite{EAB23}). The syntax is:
	\begin{center}
$P \: :: = \: \nil \mid a \, . \, P \mid P + P \mid P \pco{L} P \mid P \setminus L \mid P \, / \, L \mid C$
	\end{center}
where:

	\begin{itemize}

\item $\nil$ is the terminated process.

\item $a\, .\, \_$, for $a \in \cala_\tau$, is the action prefix operator describing a process that
initially performs action $a$.

\item $\_+\_$ is the alternative composition operator expressing a nondeterministic choice between two
processes based on their initially executable actions.

\item $\_ \pco{L} \_$, for $L \subseteq \cala$, is the parallel composition operator forcing two
processes to proceed independently on every action not in $L$ and to synchronize on every action in $L$.

\item $\_\setminus L$, for $L \subseteq \cala$, is the restriction operator, which prevents the execution of
actions in $L$.

\item $\_\,/\,L$, for $L \subseteq \cala$, is the hiding operator, which turns all the executed actions in
$L$ into the unobservable action $\tau$.

\item $C$ is a process constant equipped with a defining equation of the form $C \eqdef P$, where every
constant possibly occurring in $P$ -- including $C$ itself thus allowing for recursion -- must be in the
scope of an action prefix operator.

	\end{itemize}
\noindent 
The operational semantic rules for the process language are shown in Table~\ref{tab:op_sem} and produce the
LTS $(\procs, \cala_\tau, \! \arrow{}{} \!$) where $\arrow{}{} \! \subseteq \procs \times \cala_\tau \times
\procs$, to which the bisimulation equivalences defined in the previous section are applicable.

	\begin{table}[t]

\[\begin{array}{|rc|}
\hline
\emph{Prefix} & a \, . \, P \arrow{a}{} P \\
\emph{Choice} & \inference[]{P_1 \arrow{a}{} P_1'}{P_1 + P_2 \arrow{a}{} P_1'} \qquad\qquad\qquad
\inference[]{P_2 \arrow{a}{} P_2'}{P_1 + P_2 \arrow{a}{} P_2'} \\[0.4cm]
\emph{Interleaving} & \inference[]{P_1 \arrow{a}{} P_1' \quad a \notin L}{P_1 \pco{L} P_2 \arrow{a}{} P_1'
\pco{L} P_2} \qquad\qquad\qquad \inference[]{P_2 \arrow{a}{} P_2' \quad a \notin L}{P_1 \pco{L} P_2
\arrow{a}{} P_1 \pco{L} P_2'} \\[0.4cm]
\emph{Synchronization} & \inference[]{P_1 \arrow{a}{} P_1' \quad P_2 \arrow{a}{} P_2' \quad a \in L}{P_1
\pco{L} P_2 \arrow{a}{} P_1' \pco{L} P_2'} \\[0.4cm]	    
\emph{Restriction} & \inference[]{P \arrow{a}{} P' \quad a \notin L}{P \setminus L \arrow{a}{} P' \setminus
L} \\[0.4cm]
\emph{Hiding} & \inference[]{P \arrow{a}{} P' \quad a \in L}{P \, / \, L \arrow{\tau}{} P' \, / \, L}
\qquad\qquad\qquad \inference[]{P \arrow{a}{} P' \quad a \notin L}{P \, / \, L \arrow{a}{} P' \, / \, L}
\\[0.4cm]
\emph{Constant} & \inference[]{C \eqdef P \quad P \arrow{a}{} P'}{C \arrow{a}{} P'} \\
\hline
\end{array}\]

\caption{Operational semantic rules}
\label{tab:op_sem}

	\end{table}

%
\subsection{Weak-Bisimilarity-Based Information-Flow Properties}
\label{sec:weak_bisim_properties}
%

The intuition behind noninterference in a two-level security system is that, whenever a group of agents at
the high security level performs some actions, the effect of those actions should not be visible by any
agent at the low security level. Below is a representative selection of weak-bisimilarity-based
noninterference properties -- \emph{Nondeterministic Non-Interference} (NNI) and \emph{Non-Deducibility on
Composition} (NDC) -- whose definitions and relationships are recalled from~\cite{FG01} and, as far as
P\_BNDC is concerned, from~\cite{FR06} (this last property was not considered in~\cite{EAB23}).

	\begin{defi}\label{def:weak_bisim_properties}  

Let $P \in \procs$:

		\begin{itemize} 

\item $P \in \mathrm{BSNNI} \Longleftrightarrow P \setminus \mathcal{A}_\calh \wbis{} P \, / \,
\mathcal{A}_\calh$.

\item $P \in \mathrm{BNDC} \Longleftrightarrow$ for all $Q \in \procs$ such that every $Q' \in \reach(Q)$
has only actions in $\cala_\calh$ and for all $L \subseteq \cala_\calh$, $P \setminus \cala_\calh \wbis{}
((P \pco{L} Q) \, / \, L) \setminus \cala_\calh$. 

\item $P \in \mathrm{SBSNNI} \Longleftrightarrow$ for all $P' \in \reach(P)$, $P' \in \mathrm{BSNNI}$.  

\item $P \in \mathrm{P\_BNDC} \Longleftrightarrow$ for all $P' \in \reach(P)$, $P' \in \mathrm{BNDC}$.

\item $P \in \mathrm{SBNDC} \Longleftrightarrow$ for all $P' \in \reach(P)$ and for all $P''$ for which
there exists $h \in \cala_\calh$ such that $P' \arrow{h}{} P''$, $P' \setminus \mathcal{A}_\calh \wbis{} P''
\setminus \mathcal{A}_\calh$.
\fullbox

		\end{itemize}

	\end{defi}

	\begin{thm}\label{thm:weak_bisim_taxonomy}

$\mathrm{SBNDC} \subset \mathrm{SBSNNI} = \mathrm{P\_BNDC} \subset \mathrm{BNDC} \subset \mathrm{BSNNI}$.
\fullbox

	\end{thm}
\noindent 
Historically, one of the first and most intuitive proposals has been \emph{Bisimulation-based Strong
Nondeterministic Non-Interference} (BSNNI). Basically, it is satisfied by any process $P$ that behaves the
same when its high-level actions are prevented (as modeled by $P \setminus \mathcal{A}_\calh$) or when they
are considered as hidden, unobservable actions (as modeled by $P \, / \, \mathcal{A}_\calh $). The
equivalence between these two low-level views of $P$ states that a low-level agent cannot observe the
high-level behavior of the system. For instance, in $l \, . \, \nil + h \, . \, l \, . \, \nil$ a low-level
agent that observes the execution of $l$ cannot infer anything about the execution of $h$. Indeed, $(l \, .
\, \nil + h \, . \, l \, . \, \nil) \setminus \{ h \} \wbis{} (l \, . \, \nil + h \, . \, l \, . \, \nil) \,
/ \, \{ h \}$ because the former process is isomorphic to $l \, . \, \nil$, the latter process is isomorphic
to $l \, . \, \nil + \tau \, . \, l \, . \, \nil$, and $l \, . \, \nil \wbis{} l \, . \, \nil + \tau \, . \,
l \, . \, \nil$. 

BSNNI is not powerful enough to detect information leakages that derive from the behavior of a high-level
agent interacting with the system. For instance, $l \, . \, \nil + h_1 \, . \, h_2 \, . \, l \, . \, \nil$
is BSNNI for the same reason discussed above. However, a high-level agent like $h_1 \, . \, \nil$ enables
$h_1$ and then disables $h_2$, thus turning the low-level view of the system into $l \, . \, \nil + \tau \,
. \, \nil$, which is clearly distinguishable from $l \, . \, \nil$, as only in the former the low-level
observer may not observe $l$. \linebreak To overcome such a limitation, the most obvious solution consists
of checking explicitly the interaction on any action set $L \subseteq \cala_\calh$ between the system and
every possible high-level agent. The resulting property is \emph{Bisimulation-based Non-Deducibility on
Composition} (BNDC), which features a universal quantification over $Q$ containing only high-level actions.

To circumvent the verification problems related to such a quantifier, several properties have been proposed
that are stronger than BNDC. They all express some persistency conditions, stating that the security checks
have to be extended to all the processes reachable from a secure one. Three of the most representative among
such properties are: the variant of BSNNI that requires every reachable process to satisfy BSNNI itself,
called \emph{Strong} BSNNI (SBSNNI); the variant of BNDC that requires every reachable process to satisfy
BNDC itself, called \emph{Persistent} BNDC (P\_BNDC); and \emph{Strong} BNDC (SBNDC), which requires the
low-level view of every reachable process to be the same before and after the execution of any high-level
action, meaning that the execution of high-level actions must be completely transparent to low-level agents.
We remind that P\_BNDC and SBSNNI have been proven to be equivalent in~\cite{FR06}.

%
%
\section{Use Case: DBMS Authentication -- Part I}
\label{sec:example_p1}
%
%

Consider a multi-threaded system supporting the execution of concurrent transactions operating on a
healthcare database, where only authorized users can write their data. Then, depending on a policy governed
by the database management system (DBMS), such data can be shared with a dedicated module feeding the
training set of a machine learning (ML) facility, which is responsible for building a trained model for data
analysis purposes.

On the one hand, different authentication mechanisms can be employed to identify users and ensure data
authenticity for each transaction. We address a simple password-based mechanism ($\ms{pwd}$), a more
sophisticated two-factor authentication system ($\ms{2fa}$), and finally a scheme based on single sign on
($\ms{sso}$)~\cite{Boo20}.

On the other hand, for security reasons related to sharing sensitive data with the ML module~\cite{AC19},
only data transmitted through highly secure mechanisms, i.e., $\ms{2fa}$ and $\ms{sso}$, can be used to feed
the training set. In any case, for privacy issues, users must not be aware of whether their data are
actually chosen to train the ML model or not~\cite{BFLX22}. Hence, to avoid that the use of highly secure
authentication implicitly reveals the involvement of the ML module, the DBMS internally decides not to
consider certain transactions for the training set.

\begin{figure}[t]

\begin{center}
\begin{tikzpicture}[modal, node distance = 1cm, align=center]

\node[point] (root) [label = 135: {$\ms{Auth}$\hspace{3mm}}]    {};
\node[point] (p1)   [below left = 3cm of root, label = 180: {}] {};
\node[point] (p2)   [below right = 3cm of root, label = 0: {}]  {};
\node[point] (p3)   [below = 2cm of root, label = 180: {}]      {};
\node[point] (p6)   [below left = of p1, label = 180: {}]       {};
\node[point] (p7)   [below right = of p2, label = 0: {}]        {};

\path[->] (root) edge[bend left] node[below left]                     {$h$\hspace{-3mm}}            (p1);
\path[->] (root) edge[bend right] node[below right]                   {\hspace{-3mm}$h$}            (p2);
\path[->] (root) edge node[below right]                               {\hspace{-1mm}$\tau$}         (p3);
\path[->] (p1)   edge[out=90,in=225] node[above left]                 {$l_{\ms{sso}}$\hspace{-3mm}} (root);
\path[->] (p2)   edge[out=90,in=315] node[above right]                {\hspace{-4mm}$l_{\ms{2fa}}$} (root);
\path[->] (p3)   edge[bend left] node[below]                          {$\tau$}                      (p6);
\path[->] (p3)   edge[bend right] node[below]                         {$\tau$}                      (p7);
\path[->] (p6)   edge[out=90, in=180, looseness=1.2] node[above left] {$l_{\ms{sso}}$\hspace{-2mm}} (root);
\path[->] (p7)   edge[out=90,in=0, looseness=1.2] node[above right]   {\hspace{-3mm}$l_{\ms{2fa}}$} (root);
\path[->] (root) edge[loop, looseness=50] node[above]                 {$l_{\ms{pwd}}$}              (root);

\end{tikzpicture}
\end{center}

\caption{LTS underlying $\ms{Auth}$}
\label{fig:example_lts}

	\end{figure}
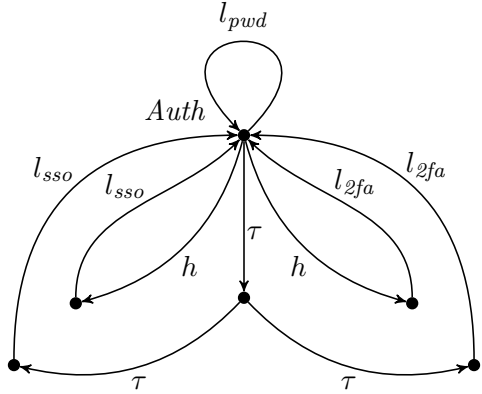

For the sake of simplicity, we concentrate on the authentication policy followed by the DBMS whenever
handling a write transaction. Therefore, we abstract away from the description of the ML module and of the
database access operations. In particular, we consider the following process term \textit{Auth}, whose LTS
is depicted in Figure~\ref{fig:example_lts}:
\cws{0}{\begin{array}{rcl}
\ms{Auth} & \!\!\! \eqdef \!\!\! & l_{\ms{pwd}} \, . \, \ms{Auth} \, + \\
& & (h \, . \, l_{\ms{sso}} \, . \, \ms{Auth} + h \, . \, l_{\ms{2fa}} \, . \, \ms{Auth}) \, + \\
& & \tau \, . \, (\tau \, . \, l_{\ms{sso}} \, . \, \ms{Auth} + \tau \, . \, l_{\ms{2fa}} \, . \, \ms{Auth})
\\
\end{array}}
The actions $l_{\star}$ express that the transaction is conducted under the authentication method
represented by $\star$. We treat them as low-level actions because they represent interactions between the
users and the DBMS. The action $h$ represents an interaction between the DBMS and the ML module, which is
deemed to be high level as the activities of the ML module must be transparent to the users.

The first subterm of $\ms{Auth}$ specifies that the DBMS is ready to offer the password-based mechanism, in
which case the transaction data will not be passed to the ML module. The second subterm models the
communication with the ML module informing that the transaction data -- which must be protected through one
of the two highly secure authentication mechanisms -- will be included in the training set. Note that in
this case the choice of the specific authentication method offered by the DBMS is nondeterministic and does
not include the password-based mechanism. The third subterm specifies that the DBMS decides internally,
through the first action $\tau$, that the transaction data will not be passed to the ML module, even if the
authentication method (chosen nondeterministically) is highly secure. Hence, in this case no interaction
with the ML module is needed. The aim of this subterm is to mimic the behavior of the second subterm, thus
acting as an obfuscation mechanism that shall not allow any user to detect the potential involvement of the
ML module by simply observing the used authentication method. 

Formally, the success of this obfuscation is guaranteed if the interaction with the ML module does not
interfere with the low-level view of the system observed by any user, which can be verified as a
noninterference property. More specifically, the ML module represents the high-level portion of the system
that is expected not to interfere with the low-level behavior of any user interacting with the DBMS, thus
justifying the use of the high-level action $h$ modeling the interaction between such a module and the DBMS.

\begin{figure}[t]

\begin{center}
\begin{tikzpicture}[modal, node distance = 1cm, align=center]

\node[point] (root') [label = 135: {$s_1$\hspace{3mm}}]               {};
\node[]      (p1')   [below left = 2.83cm of root', label = 270: {}]  {};
\node[]      (p2')   [below right = 2.83cm of root', label = 270: {}] {};
\node[point] (p3')   [below = 2cm of root', label = 180: {$s_2$}]     {};
\node[point] (p6')   [below left = of p1', label = 270: {$s_3$}]      {};
\node[point] (p7')   [below right = of p2', label = 270: {$s_4$}]     {};

\path[->] (root') edge node[below right]                              {\hspace{-1mm}$\tau$}         (p3');
\path[->] (p3')   edge[bend left] node[below]                         {$\tau$}                      (p6');
\path[->] (p3')   edge[bend right] node[below]                        {$\tau$}                      (p7');
\path[->] (p6')   edge[out=90,in=180,looseness=1.2] node[above left]  {$l_{\ms{sso}}$\hspace{-2mm}} (root');
\path[->] (p7')   edge[out=90,in=0,looseness=1.2] node[above right]   {\hspace{-3mm}$l_{\ms{2fa}}$} (root');
\path[->] (root') edge[loop,looseness=50] node[above]                 {$l_{\ms{pwd}}$}              (root');

\node[point] (root) [label = 135: {$r_1$\hspace{3mm}}, right = 7.5cm of root'] {};
\node[point] (p1)   [below left = 3cm of root, label = 270: {$r'_3$}]          {};
\node[point] (p2)   [below right = 3cm of root, label = 270: {$r'_4$}]         {};
\node[point] (p3)   [below = 2cm of root, label = 180: {$r_2$}]                {};
\node[point] (p6)   [below left = of p1, label = 270: {$r_3$}]                 {};
\node[point] (p7)   [below right = of p2, label = 270: {$r_4$}]                {};

\path[->] (root) edge[bend left]  node[below left]     {\textcolor{magenta}{$\tau$}\hspace{-4mm}}  (p1);
\path[->] (root) edge[bend right] node[below right]    {\hspace{-4mm}\textcolor{magenta}{$\tau$}}  (p2);
\path[->] (root) edge node[below right]                              {\hspace{-1mm}$\tau$}         (p3);
\path[->] (p1)   edge[out=90,in=225] node[above left]                {$l_{\ms{sso}}$\hspace{-3mm}} (root);
\path[->] (p2)   edge[out=90,in=315] node[above right]               {\hspace{-4mm}$l_{\ms{2fa}}$} (root);
\path[->] (p3)   edge[bend left]  node[below]                        {$\tau$}                      (p6);
\path[->] (p3)   edge[bend right] node[below]                        {$\tau$}                      (p7);
\path[->] (p6)   edge[out=90,in=180,looseness=1.2] node[above left]  {$l_{\ms{sso}}$\hspace{-2mm}} (root);
\path[->] (p7)   edge[out=90,in=0,looseness=1.2] node[above right]   {\hspace{-3mm}$l_{\ms{2fa}}$} (root);
\path[->] (root) edge[loop,looseness=50] node[above]                 {$l_{\ms{pwd}}$}              (root);

\end{tikzpicture}
\end{center}

\caption{LTSs of the low-level views of $\ms{Auth}$: $\ms{Auth} \setminus \cala_\calh$ (left) and $\ms{Auth}
\, / \, \cala_\calh$ (right)}
\label{fig:example_low_views}

	\end{figure}
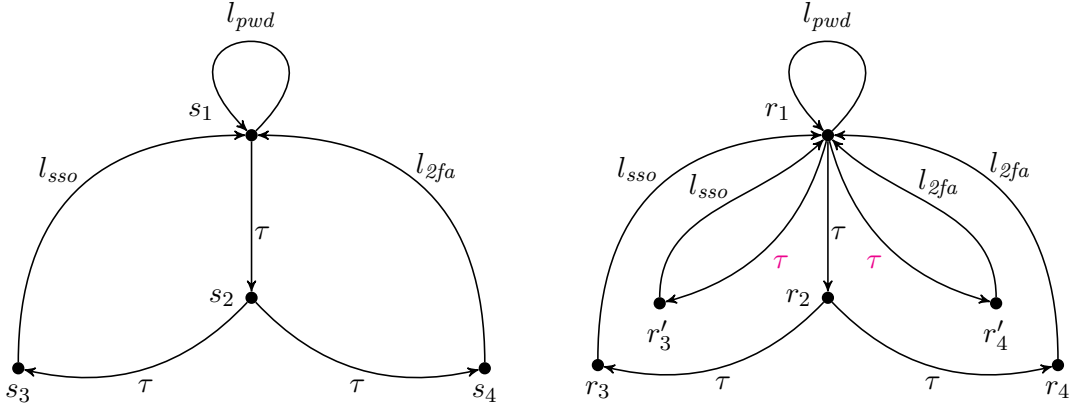

As far as $\wbis{}$-based noninterference is concerned, $\ms{Auth}$ does not leak any information from the
high level to the low level. More precisely, the system is SBSNNI, and hence also BNDC and BSNNI by virtue
of Theorem~\ref{thm:weak_bisim_taxonomy}. Indeed, by observing Figure~\ref{fig:example_low_views} -- where
the $h$-actions are forbidden on the left while they are transformed into the colored $\tau$-actions on the
right -- it is easy to see that $\ms{Auth}$ is BSNNI, i.e., $\ms{Auth} \setminus \cala_\calh \wbis{}
\ms{Auth} \, / \, \cala_\calh$. The weak bisimulation relating the two low-level views of $\ms{Auth}$ is
given by the following partition of the disjoint union of the two state spaces: \cws{0}{\{\{ s_1, r_1 \}, \{
s_2, r_2 \}, \{ s_3, r_3, r_3' \}, \{ s_4, r_4, r_4' \}\}} Since the only high-level action is enabled at
the initial state of $\ms{Auth}$, it then follows that $\ms{Auth}$ is SBSNNI as well.

%
%
\section{Security Properties Based on Branching Bisimilarity}
\label{sec:branching_properties} 
%
%

While the literature on noninterference mainly concentrates on weak bisimulation semantics, in this article
we address information-flow security in terms of branching bisimilarity.

	\begin{defi}\label{def:branching_bisim_properties}  

BrSNNI, BrNDC, SBrSNNI, P\_BrNDC, and SBrNDC are obtained from the corresponding properties in
Definition~\ref{def:weak_bisim_properties} by replacing the weak bisimilarity check ($\wbis{}$) with the
branching bisimilarity check ($\wbis{\rm b}$).
\fullbox

	\end{defi}

\noindent
In this section we first study their preservation and compositionality characteristics so as to assess their
usefulness (Section~\ref{sec:branching_pres_comp}) and then we investigate the relationships among them and
with the corresponding properties based on weak bisimilarity (Section~\ref{sec:branching_taxonomy}).

%
\subsection{Preservation and Compositionality}
\label{sec:branching_pres_comp} 
%

Similar to the weak bisimilarity case~\cite{FG01}, all the $\wbis{\rm b}$-based noninterference properties
turn out to be preserved by $\wbis{\rm b}$. This means that, whenever a process $P_1$ is secure under any of
such properties, then every other branching bisimilar process $P_2$ is secure too according to the same
property. This is very useful for automated property verification, as it allows one to work with the process
with the smallest state space among the equivalent ones. To establish preservation, we first have to prove
that $\wbis{\rm b}$ is a congruence with respect to $\_ \setminus L$, $\_ \, / \, L$, and $\_ \pco{L} \_$
for an arbitrary $L \subseteq \cala$, because these three operators were not considered in the congruence
results of~\cite{GW96,Gla93}.

	\begin{lem}\label{lem:congr_restr_hiding_parallel}

Let $P_1, P_2 \in \procs$. If $P_1 \wbis{\rm b} P_2$ then:

		\begin{itemize}

\item $P_1 \setminus L \wbis{\rm b} P_2 \setminus L$ for all $L \subseteq \cala$.

\item $P_1 \, / \, L \wbis{\rm b} P_2 \, / \, L$ for all $L \subseteq \cala$.

\item $P_1 \pco{L} P \wbis{\rm b} P_2 \pco{L} P$ and $P \pco{L} P_1 \wbis{\rm b} P \pco{L} P_2$ for all $L
\subseteq \cala$ and $P \in \procs$.

		\end{itemize}

		\begin{proof}
Let $\calb$ be a branching bisimulation containing the pairs $(P_1, P_2)$ and $(P_2, P_1)$ and consider an
arbitrary set $L \subseteq \cala$ and an arbitrary process $P \in \procs$. We have that:

			\begin{itemize}

\item $P_1 \setminus L \wbis{\rm b} P_2 \setminus L$ because the symmetric relation $\calb' = \{ (Q_1
\setminus L, Q_2 \setminus L) \mid (Q_1, Q_2) \in \calb \}$ \linebreak is a branching bisimulation too, as
we now show via the following two cases based on the operational semantic rules in Table~\ref{tab:op_sem}:

				\begin{itemize}

\item If $Q_1 \setminus L \arrow{\tau}{} Q'_1 \setminus L$ with $Q_1 \arrow{\tau}{} Q'_1$, then either
$(Q'_1, Q_2) \in \calb$, or there exist $\bar{Q}_2$ and $Q'_2$ such that $Q_2 \warrow{\tau^{*}}{} \bar{Q}_2
\arrow{\tau}{} Q'_2$ with $(Q_1, \bar{Q}_2) \in \calb$ and $(Q'_1, Q'_2) \in \calb$. Since the restriction
operator does not apply to $\tau$, in the former subcase $Q_2 \setminus L$ is allowed to stay idle with
$(Q'_1 \setminus L, Q_2 \setminus L) \in \calb'$, while in the latter subcase $Q_2 \setminus L
\warrow{\tau^{*}}{} \bar{Q}_2 \setminus L \arrow{\tau}{} Q'_2 \setminus L$, with $(Q_1 \setminus L,
\bar{Q}_2 \setminus L) \in \calb'$ and $(Q'_1 \setminus L, Q'_2 \setminus L) \in \calb'$.

\item If $Q_1 \setminus L \arrow{a}{} Q'_1 \setminus L$ with $Q_1 \arrow{a}{} Q'_1$ and $a \notin L \cup \{
\tau \}$, then there exist $\bar{Q}_2$ and $Q'_2$ such that $Q_2 \warrow{\tau^{*}}{} \bar{Q}_2 \arrow{a}{}
Q'_2$ with $(Q_1, \bar{Q}_2) \in \calb$ and $(Q'_1, Q'_2) \in \calb$. Since the restriction operator does
not apply to $\tau$ and $a \notin L$, it follows that $Q_2 \setminus L \warrow{\tau^{*}}{} \bar{Q}_2
\setminus L \arrow{a}{} Q'_2 \setminus L$ with $(Q_1 \setminus L, \bar{Q}_2 \setminus L) \in \calb'$ and
$(Q'_1 \setminus L, Q'_2 \setminus L) \in \calb'$.

				\end{itemize}

\item $P_1 \, / \, L \wbis{\rm b} P_2 \, / \, L$ because the symmetric relation $\calb' = \{ (Q_1 \, / \, L,
Q_2 \, / \, L) \mid (Q_1, Q_2) \in \calb \}$ \linebreak is a branching bisimulation too, as we now show via
the following two cases based on the operational semantic rules in Table~\ref{tab:op_sem}:

				\begin{itemize}

\item If $Q_1 \, / \, L \arrow{\tau}{} Q'_1 \, / \, L$ with $Q_1 \arrow{\tau}{} Q'_1$, then either $(Q'_1,
Q_2) \in \calb$, or there exist $\bar{Q}_2$ and~$Q'_2$ such that $Q_2 \warrow{\tau^{*}}{} \bar{Q}_2
\arrow{\tau}{} Q'_2$ with $(Q_1, \bar{Q}_2) \in \calb$ and $(Q'_1, Q'_2) \in \calb$. Since the hiding
operator does not apply to $\tau$, in the former subcase $Q_2 \, / \, L$ is allowed to stay idle with $(Q'_1
\, / \, L, Q_2 \, / \, L) \in \calb'$, while in the latter subcase $Q_2 \, / \, L \warrow{\tau^{*}}{}
\bar{Q}_2 \, / \, L \arrow{\tau}{} Q'_2 \, / \, L$ with $(Q_1 \, / \, L , \bar{Q}_2 \, / \, L) \in \calb'$
and $(Q'_1 \, / \, L, Q'_2 \, / \, L) \in \calb'$.

\item If $Q_1 \, / \, L \arrow{a}{} Q'_1 \, / \, L$ with $Q_1 \arrow{b}{} Q'_1$ and $b \in L \land a = \tau$
or $b \notin L \cup \{ \tau \} \land a = b$, then there exist $\bar{Q}_2$ and $Q'_2$ such that $Q_2
\warrow{\tau^{*}}{} \bar{Q}_2 \arrow{b}{} Q'_2$ with $(Q_1, \bar{Q}_2) \in \calb$ and $(Q'_1, Q'_2) \in
\calb$. Since the hiding operator does not apply to $\tau$, it follows that $Q_2 \, / \, L
\warrow{\tau^{*}}{} \bar{Q}_2 \, / \, L \arrow{a}{} Q'_2 \, / \, L$, with $(Q_1 \, / \, L , \bar{Q}_2 \, /
\, L) \in \calb'$ and $(Q'_1 \, / \, L, Q'_2 \, / \, L) \in \calb'$. 

				\end{itemize}

\item $P_1 \pco{L} P \wbis{\rm b} P_2 \pco{L} P$ because the symmetric relation $\calb' = \{ (Q_1 \pco{L} Q,
Q_2 \pco{L} Q) \mid (Q_1, Q_2) \in \calb \land Q \in \procs \}$ is a branching bisimulation too, as we now
show via the following three cases based on the operational semantic rules in Table~\ref{tab:op_sem}:

				\begin{itemize}

\item If $Q_1 \pco{L} Q \arrow{a}{} Q'_1 \pco{L} Q$ with $Q_1 \arrow{a}{} Q'_1$ and $a \notin L$, then
either $a = \tau$ and $(Q'_1, Q_2) \in \calb$, or there exist $\bar{Q}_2$ and $Q'_2$ such that $Q_2
\warrow{\tau^{*}}{} \bar{Q}_2 \arrow{a}{} Q'_2$ with $(Q_1, \bar{Q}_2) \in \calb$ and $(Q'_1, Q'_2) \in
\calb$. In the former subcase $Q_2 \pco{L} Q$ is allowed to stay idle with $(Q'_1 \pco{L} Q, Q_2 \pco{L} Q)
\in \calb'$, while in the latter subcase $Q_2 \pco{L} Q \warrow{\tau^{*}}{} \bar{Q}_2 \pco{L} Q \arrow{a}{}
Q'_2 \pco{L} Q$ with $(Q_1 \pco{L} Q, \bar{Q}_2 \pco{L} Q) \in \calb'$ and $(Q'_1 \pco{L} Q, Q'_2 \pco{L} Q)
\in \calb'$.

\item The case $Q_1 \pco{L} Q \arrow{a}{} Q_1 \pco{L} Q'$ with $Q \arrow{a}{} Q'$ and $a \notin L$ is
trivial.

\item If $Q_1 \pco{L} Q \arrow{a}{} Q'_1 \pco{L} Q'$ with $Q_1 \arrow{a}{} Q'_1$, $Q \arrow{a}{} Q'$, and $a
\in L$, then there exist $\bar{Q}_2$ and $Q'_2$ such that $Q_2 \warrow{\tau^{*}}{} \bar{Q}_2 \arrow{a}{}
Q'_2$ with $(Q_1, \bar{Q}_2) \in \calb$ and $(Q'_1, Q'_2) \in \calb$. Thus $Q_2 \pco{L} Q \linebreak
\warrow{\tau^{*}}{} \bar{Q}_2 \pco{L} Q \arrow{a}{} Q'_2 \pco{L} Q'$ with $(Q_1 \pco{L} Q, \bar{Q}_2\pco{L}
Q) \in \calb'$ and $(Q'_1 \pco{L} Q', Q'_2 \pco{L} Q') \in \calb'$.

				\end{itemize}

\noindent
The proof of $P \pco{L} P_1 \wbis{\rm b} P \pco{L} P_2$ is similar.
\qedhere

			\end{itemize}

		\end{proof}

	\end{lem}

	\begin{thm}\label{thm:preservation}

Let $P_1, P_2 \in \procs$ and $\calp \in \{ \mathrm{BrSNNI}, \mathrm{BrNDC}, \mathrm{SBrSNNI},
\mathrm{P\_BrNDC}, \mathrm{SBrNDC}\}$. If $P_1 \wbis{\rm b} P_2$, then $P_1 \in \calp \Longleftrightarrow
P_2 \in \calp$.

		\begin{proof}
A straightforward consequence of the definition of the five properties, i.e.,
Definition~\ref{def:branching_bisim_properties}, and Lemma~\ref{lem:congr_restr_hiding_parallel}.
\qedhere

		\end{proof}

	\end{thm}
\noindent 
As far as modular verification is concerned, like in the weak-bisimilarity-based case~\cite{FG01} only the
local properties SBrSNNI and SBrNDC are compositional, i.e., are preserved by some operators of the calculus
in certain circumstances; this holds also for P\_BrNDC because we will see later on that P\_BrNDC coincides
with SBrSNNI (Theorem~\ref{thm:branching_taxonomy_1}). Unlike the compositionality results presented
in~\cite{FG01}, ours are related not only to parallel composition and restriction, but also to action prefix
and hiding. Like in the weak-bisimilarity-based case~\cite{FG01}, no property relying on branching
bisimilarity is compositional with respect to alternative composition. For instance, let us consider
processes $P_1$ and $P_2$ respectively given by $l \, . \, \nil$ and $h \, . \, \nil$. Both are BrSNNI, as
$l \, . \, \nil \setminus \{ h \} \wbis{\rm b} l \, . \, \nil \, / \, \{ h \}$ and $h \, . \, \nil \setminus
\{ h \} \wbis{\rm b} h \, . \, \nil \, / \, \{ h \}$, but $P_1 + P_2 \notin \mathrm{BrSNNI}$ because $(l \,
. \, \nil + h \, . \, \nil) \setminus \{ h \} \wbis{\rm b} l \, . \, \nil \not\wbis{\rm b} l \, . \, \nil +
\tau \, . \, \nil \wbis{\rm b} (l \, . \, \nil + h \, . \, \nil) \, / \, \{ h \}$. It can be easily checked
that $P_1 + P_2 \notin \calp$ also for $\calp \in \{ \mathrm{BrNDC}, \mathrm{SBrSNNI}, {\mathrm{P\_BrNDC}},
\mathrm{SBrNDC} \}$.  

Compositionality with respect to parallel composition $\pco{L}$ is limited, for SBrSNNI and P\_BrNDC, to the
case in which no synchronization can take place between high-level actions, i.e., $L \subseteq \cala_\call$,
while in~\cite{FG01} the compositionality of SBSNNI holds for every $L \subseteq \cala$. As an example,
$P_1$ given by $h \, . \, \nil + l_1 \, . \, \nil + \tau \, . \, \nil$ and $P_2$ given by $h \, . \, \nil +
l_2 \, . \, \nil + \tau \, . \, \nil$ are SBrSNNI, but $P_1 \pco{\{ h \}} P_2$ is not because the transition
$(P_1 \pco{\{ h \}} P_2) \, / \, \cala_\calh \arrow{\tau}{} (\nil \pco{\{ h \}} \nil) \, / \, \cala_\calh$
arising from the synchronization between the two $h$-actions cannot be matched by $(P_1 \pco{\{ h \}} P_2)
\setminus \cala_\calh$ in the branching bisimulation game. As a matter of fact, the only two possibilities
are $(P_1 \pco{\{ h \}} P_2) \setminus \cala_\calh \warrow{\tau^{*}}{} (P_1 \pco{\{ h \}} P_2) \setminus
\cala_\calh \arrow{\tau}{} (\nil \pco{\{ h \}} P_2) \setminus \cala_\calh \arrow{\tau}{} (\nil \pco{\{ h \}}
\nil) \setminus \cala_\calh$ as well as $(P_1 \pco{\{ h \}} P_2) \setminus \cala_\calh \warrow{\tau^{*}}{}
(P_1 \pco{\{ h \}} P_2) \setminus \cala_\calh \arrow{\tau}{} (P_1 \pco{\{ h \}} \nil) \setminus \cala_\calh
\arrow{\tau}{} (\nil \pco{\{ h \}} \nil) \setminus \cala_\calh$ but neither $(\nil \pco{\{ h \}} P_2)
\setminus \cala_\calh$ nor $(P_1 \pco{\{ h \}} \nil) \setminus \cala_\calh$ is branching bisimilar to $(P_1
\pco{\{ h \}} P_2) \setminus \cala_\calh$ when $l_1 \neq l_2$. Note that $(P_1 \pco{\{ h \}} P_2) \, / \,
\cala_\calh \wbis{} (P_1 \pco{\{ h \}} P_2) \setminus \cala_\calh$ because $(P_1 \pco{\{ h \}} P_2) \, / \,
\cala_\calh \arrow{\tau}{} (\nil \pco{\{ h \}} \nil) \, / \, \cala_\calh$ is matched by $(P_1 \pco{\{ h \}}
P_2) \setminus \cala_\calh \warrow{\tau^{*}}{} (\nil \pco{\{ h \}} \nil) \setminus \cala_\calh$. However, it
is not only a matter of the higher discriminating power of $\wbis{\rm b}$ with respect to $\wbis{}$. If we
used the CCS parallel composition operator~\cite{Mil89a}, which turns into $\tau$ the synchronization of two
actions thus combining communication with hiding, then the parallel composition of $P_1$ and $P_2$ with
restriction on $\cala_\calh$ would be able to respond, in the branching bisimulation game, with a single
$\tau$-transition reaching the parallel composition of $\nil$ and $\nil$ with restriction on $\cala_\calh$.

To establish compositionality, we first prove some ancillary results about parallel composition,
restriction, and hiding under SBrSNNI and SBrNDC.

	\begin{lem}\label{lem:compositionality}

Let $P_1, P_2, P \in \procs$. Then:

		\begin{enumerate}

\item If $P_1, P_2 \in \mathrm{SBrSNNI}$ and $L \subseteq \cala_\call$, then $(Q_1 \pco{L} Q_2) \setminus
\cala_\calh \wbis{\rm b} (R_1 \pco{L} R_2) \, / \, \cala_\calh$ \linebreak for all $Q_1, R_1 \in
\reach(P_1)$ and $Q_2, R_2 \in \reach(P_2)$ such that $Q_1 \pco{L} Q_2, R_1 \pco{L} R_2 \in \linebreak
\reach(P_{1} \pco{L} P_{2})$, $Q_1 \setminus \cala_\calh \wbis{\rm b} R_1 \, / \, \cala_\calh$, and $Q_2
\setminus \cala_\calh \wbis{\rm b} R_2 \, / \, \cala_{\calh}$.

\item If $P \in \mathrm{SBrSNNI}$ and $L \subseteq \cala$, then $(Q \, / \, \cala_\calh) \setminus L
\wbis{\rm b} (R \setminus L) \, / \, \cala_\calh$ for all $Q, R \in \reach(P)$ such that $Q \, / \,
\cala_\calh \wbis{\rm b} R \setminus \cala_\calh$.

\item If $P_1, P_2 \in \mathrm{SBrNDC}$ and $L \subseteq \cala$, then $(Q_1 \pco{L} Q_2) \setminus
\cala_\calh \wbis{\rm b} (R_1 \pco{L} R_2) \setminus \cala_\calh$ \linebreak for all $Q_1, R_1 \in
\reach(P_1)$ and $Q_2, R_2 \in \reach(P_2)$ such that $Q_1 \pco{L} Q_2, R_1 \pco{L} R_2 \in \reach(P_1
\pco{L} P_2)$, $Q_1 \setminus \cala_\calh \wbis{\rm b} R_1 \setminus \cala_\calh$ and $Q_2 \setminus
\cala_\calh \wbis{\rm b} R_2 \setminus \cala_\calh$.

		\end{enumerate}

		\begin{proof}
Let $\calb$ be a symmetric relation containing all the pairs of processes that have to be shown to be
branching bisimilar according to the property considered among the three stated above:

			\begin{enumerate}

\item Starting from $(Q_1 \pco{L} Q_2) \setminus \cala_\calh$ and $(R_1 \pco{L} R_2) \, / \, \cala_\calh$
related by $\calb$, so that $Q_1 \setminus \cala_\calh \wbis{\rm b} R_1 \, / \, \cala_\calh$ and $Q_2
\setminus \cala_\calh \wbis{\rm b} R_2 \, / \, \cala_{\calh}$, in the branching bisimulation game there are
twelve cases based on the operational semantic rules in Table~\ref{tab:op_sem}. In the first five cases, it
is $(Q_1 \pco{L} Q_2) \setminus \cala_\calh$ to move first:

				\begin{itemize}

\item If $(Q_1 \pco{L} Q_2) \setminus \cala_\calh \arrow{l}{} (Q'_1 \pco{L} Q_2) \setminus \cala_\calh$ with
$Q_1 \arrow{l}{} Q'_1$ and $l \notin L$, then \linebreak $Q_1 \setminus \cala_\calh \arrow{l}{} Q'_1
\setminus \cala_\calh$ as $l \notin \cala_\calh$. From $Q_1 \setminus \cala_\calh \wbis{\rm b} R_1 \, / \,
\cala_\calh$ it follows that there exist $\bar{R}_1$ and $R'_1$ such that $R_1 \, / \, \cala_\calh
\warrow{\tau^{*}}{} \bar{R}_1 \, / \, \cala_\calh \arrow{l}{} R'_1 \, / \, \cala_\calh$ with $Q_1 \setminus
\cala_\calh \wbis{\rm b} \bar{R}_1 \, / \, \cala_\calh$ and $Q'_1 \setminus \cala_\calh \wbis{\rm b} R'_1 \,
/ \, \cala_\calh$. Since synchronization does not apply to $\tau$ and~$l$, it follows that $(R_1 \pco{L}
R_2) \, / \, \cala_\calh \warrow{\tau^{*}}{} (\bar{R}_1 \pco{L} R_2) \, / \, \cala_\calh \arrow{l}{} (R'_1
\pco{L} R_2) \, / \, \cala_\calh$ with $((Q_1 \pco{L} Q_2) \setminus \cala_\calh, (\bar{R}_1 \pco{L} R_2) \,
/ \, \cala_\calh) \in \calb$ and $((Q'_1 \pco{L} Q_2) \setminus \cala_\calh, (R'_1 \pco{L} R_2) \, / \,
\cala_\calh) \in \calb$.

\item If $(Q_1 \pco{L} Q_2) \setminus \cala_\calh \arrow{l}{} (Q_1 \pco{L} Q'_2) \setminus \cala_\calh$ with
$Q_2 \arrow{l}{} Q'_2$ and $l \notin L$, then the proof is similar to the one of the previous case.

\item If $(Q_1 \pco{L} Q_2) \setminus \cala_\calh \arrow{l}{} (Q'_1 \pco{L} Q'_2) \setminus \cala_\calh$
with $Q_i \arrow{l}{} Q'_i$ for $i \in \{ 1, 2 \}$ and $l \in L$, then $Q_i \setminus \cala_\calh
\arrow{l}{} Q'_i \setminus \cala_\calh$ as $l \notin \cala_\calh$. From $Q_i \setminus \cala_\calh \wbis{\rm
b} R_i \, / \, \cala_\calh$ it follows that there exist $\bar{R}_i$ and $R'_i$ such that $R_i \, / \,
\cala_\calh \warrow{\tau^{*}}{} \bar{R}_i \, / \, \cala_\calh \arrow{l}{} R'_i \, / \, \cala_\calh$ with
$Q_i \setminus \cala_\calh \wbis{\rm b} \bar{R}_i \, / \, \cala_\calh$ and $Q'_i \setminus \cala_\calh
\wbis{\rm b} R'_i \, / \, \cala_\calh$. Since synchronization does not apply to $\tau$, it follows that
$(R_1 \pco{L} R_2) \, / \, \cala_\calh \warrow{\tau^{*}}{} (\bar{R}_1 \pco{L} \bar{R}_2) \, / \, \cala_\calh
\arrow{l}{} (R'_1 \pco{L} R'_2) \, / \, \cala_\calh$ with $((Q_1 \pco{L} Q_2) \setminus \cala_\calh,
\linebreak (\bar{R}_1 \pco{L} \bar{R}_2) \, / \, \cala_\calh) \in \calb$ and $((Q'_1 \pco{L} Q'_2) \setminus
\cala_\calh, (R'_1 \pco{L} R'_2) \, / \, \cala_\calh) \in \calb$.

\item If $(Q_1 \pco{L} Q_2) \setminus \cala_\calh \arrow{\tau}{} (Q'_1 \pco{L} Q_2) \setminus \cala_\calh$
with $Q_1 \arrow{\tau}{} Q'_1$, then $Q_1 \setminus \cala_\calh \arrow{\tau}{} Q'_1 \setminus \cala_\calh$
as $\tau \notin \cala_\calh$. From $Q_1 \setminus \cala_\calh \wbis{\rm b} R_1 \, / \, \cala_\calh$ it
follows that either $Q'_1 \setminus \cala_\calh \wbis{\rm b} R_1 \, / \, \cala_\calh$, or there exist
$\bar{R}_1$ and $R'_1$ such that $R_1 \, / \, \cala_\calh \warrow{\tau^{*}}{} \bar{R}_1 \, / \, \cala_\calh
\arrow{\tau}{} R'_1 \, / \, \cala_\calh$ with $Q_1 \setminus \cala_\calh \wbis{\rm b} \bar{R}_1 \, / \,
\cala_\calh$ and $Q'_1 \setminus \cala_\calh \wbis{\rm b} R'_1 \, / \, \cala_\calh$. In the former subcase
$(R_1 \pco{L} R_2) \, / \, \cala_\calh$ is allowed to stay idle with $((Q'_1 \pco{L} Q_2) \setminus
\cala_\calh, (R_1 \pco{L} R_2) \, / \, \cala_\calh) \in \calb$, while in the latter subcase, since
synchronization does not apply to $\tau$, it follows that $(R_1 \pco{L} R_2) \, / \, \cala_\calh
\warrow{\tau^{*}}{} (\bar{R}_1 \pco{L} R_2) \, / \, \cala_\calh \arrow{\tau}{}$ \linebreak $(R'_1 \pco{L}
R_2) \, / \, \cala_\calh$ with $((Q_1 \pco{L} Q_2) \setminus \cala_\calh, (\bar{R}_1 \pco{L} R_2) \, / \,
\cala_\calh) \in \calb$ and $((Q'_1 \pco{L} Q_2) \setminus \cala_\calh, \linebreak (R'_1 \pco{L} R_2) \, /
\, \cala_\calh) \in \calb$.

\item If $(Q_1 \pco{L} Q_2) \setminus \cala_\calh \arrow{\tau}{} (Q_1 \pco{L} Q'_2)\setminus \cala_\calh$
with $Q_2 \arrow{\tau}{} Q'_2$, then the proof is similar to the one of the previous case. 

				\end{itemize}

\noindent
In the other seven cases, instead, it is $(R_1 \pco{L} R_2) \, / \, \cala_\calh$ to move first:

				\begin{itemize}

\item If $(R_1 \pco{L} R_2) \, / \, \cala_\calh \arrow{l}{} (R'_1 \pco{L} R_2) \, / \, \cala_\calh$ with
$R_1 \arrow{l}{} R'_1$ and $l \notin L$, then $R_1 \, / \, \cala_\calh \arrow{l}{}$ \linebreak $R'_1 \, / \,
\cala_\calh$ as $l \notin \cala_\calh$. From $R_1 \, / \, \cala_\calh \wbis{\rm b} Q_1 \setminus
\cala_\calh$ it follows that there exist $\bar{Q}_1$ and $Q'_1$ such that $Q_1 \setminus \cala_\calh
\warrow{\tau^{*}}{} \bar{Q}_1 \setminus \cala_\calh \arrow{l}{} Q'_1 \setminus \cala_\calh$ with $R_1 \, /
\, \cala_\calh \wbis{\rm b} \bar{Q}_1 \setminus \cala_\calh$ and $R'_1 \, / \, \cala_\calh \wbis{\rm b} Q'_1
\setminus \cala_\calh$. Since synchronization does not apply to $\tau$ and~$l$, it follows that $(Q_1
\pco{L} Q_2) \setminus \cala_\calh \linebreak \warrow{\tau^{*}}{} (\bar{Q}_1 \pco{L} Q_2) \setminus
\cala_\calh \arrow{l}{} (Q'_1 \pco{L} Q_2) \setminus \cala_\calh$ with $((R_1 \pco{L} R_2) / \cala_\calh,
(\bar{Q}_1 \pco{L} Q_2) \setminus \cala_\calh) \in \calb$ and $((R'_1 \pco{L} R_2) / \cala_\calh, (Q'_1
\pco{L} Q_2) \setminus \cala_\calh) \in \calb$.

\item If $(R_1 \pco{L} R_2) \, / \, \cala_\calh \arrow{l}{} (R_1 \pco{L} R'_2) \, / \, \cala_\calh$ with
$R_2 \arrow{l}{} R'_2$ and $l \notin L$, then the proof is similar to the one of the previous case. 

\item If $(R_1 \pco{L} R_2) \, / \, \cala_\calh \arrow{l}{} (R'_1 \pco{L} R'_2) \, / \, \cala_\calh$ with
$R_i \arrow{l}{} R'_i$ for $i \in \{ 1, 2 \}$ and $l \in L$, then $R_i \, / \, \cala_\calh \arrow{l}{} R'_i
\, / \, \cala_\calh$ as $l \notin \cala_\calh$. From $R_i \, / \, \cala_\calh \wbis{\rm b} Q_i \setminus
\cala_\calh$ it follows that there exist $\bar{Q}_i$ and $Q'_i$ such that $Q_i \setminus \cala_\calh
\warrow{\tau^{*}}{} \bar{Q}_i \setminus \cala_\calh \arrow{l}{} Q'_i \setminus \cala_\calh$ with $R_i \, /
\, \cala_\calh \wbis{\rm b} \bar{Q}_i \setminus \cala_\calh$ and $R'_i \, / \, \cala_\calh \wbis{\rm b} Q'_i
\setminus \cala_\calh$. Since synchronization does not apply to $\tau$, it follows that $(Q_1 \pco{L} Q_2)
\setminus \cala_\calh \warrow{\tau^{*}}{} (\bar{Q}_1 \pco{L} \bar{Q}_2) \setminus \cala_\calh \arrow{l}{}
(Q'_1 \pco{L} Q'_2) \setminus \cala_\calh$ with $((R_1 \pco{L} R_2) / \cala_\calh, \linebreak (\bar{Q}_1
\pco{L} \bar{Q}_2) \setminus \cala_\calh) \in \calb$ and $((R'_1 \pco{L} R'_2) / \cala_\calh, (Q'_1 \pco{L}
Q'_2) \setminus \cala_\calh) \in \calb$.

\item If $(R_1 \pco{L} R_2) \, / \, \cala_\calh \arrow{\tau}{} (R'_1 \pco{L} R_2) \, / \, \cala_\calh$ with
$R_1 \arrow{\tau}{} R'_1$, then $R_1 \, / \, \cala_\calh \arrow{\tau}{} R'_1 \, / \, \cala_\calh$ as $\tau
\notin \cala_\calh$. From $R_1 \, / \, \cala_\calh \wbis{\rm b} Q_1 \setminus \cala_\calh$ it follows that
either $R'_1 \, / \, \cala_\calh \wbis{\rm b} Q_1 \setminus \cala_\calh$, or there exist $\bar{Q}_1$ and
$Q'_1$ such that $Q_1 \setminus \cala_\calh \warrow{\tau^{*}}{} \bar{Q}_1 \setminus \cala_\calh
\arrow{\tau}{} Q'_1 \setminus \cala_\calh$ with $R_1 \, / \, \cala_\calh \wbis{\rm b} \bar{Q}_1 \setminus
\cala_\calh$ and $R'_1 \, / \, \cala_\calh \wbis{\rm b} Q'_1 \setminus \cala_\calh$. In the former subcase
$(Q_1 \pco{L} Q_2) \setminus \cala_\calh$ is allowed to stay idle with $((R'_1 \pco{L} R_2) / \cala_\calh,
(Q_1 \pco{L} Q_2) \setminus \cala_\calh) \in \calb$, while in the latter subcase, since synchronization does
not apply to $\tau$, it follows that $(Q_1 \pco{L} Q_2) \setminus \cala_\calh \linebreak \warrow{\tau^{*}}{}
(\bar{Q}_1 \pco{L} Q_2) \setminus \cala_\calh \arrow{\tau}{} (Q'_1 \pco{L} Q_2) \setminus \cala_\calh$ with
$((R_1 \pco{L} R_2) / \cala_\calh, (\bar{Q}_1 \pco{L} Q_2) \setminus \cala_\calh) \in \calb$ and $((R'_1
\pco{L} R_2) / \cala_\calh, (Q'_1 \pco{L} Q_2) \setminus \cala_\calh) \in \calb$.

\item If $(R_1 \pco{L} R_2) \, / \, \cala_\calh \arrow{\tau}{} (R_1 \pco{L} R'_2) \, / \, \cala_\calh$ with
$R_2 \arrow{\tau}{} R'_2$, then the proof is similar to the one of the previous case.

\item If $(R_1 \pco{L} R_2) \, / \, \cala_\calh \arrow{\tau}{} (R'_1 \pco{L} R_2) \, / \, \cala_\calh$ with
$R_1 \arrow{h}{} R'_1$ and $h \notin L$, then $R_1 \, / \, \cala_\calh \arrow{\tau}{}$ \linebreak $R'_1 \, /
\, \cala_\calh$ as $h \in \cala_\calh$. From $R_1 \, / \, \cala_\calh \wbis{\rm b} Q_1 \setminus
\cala_\calh$ it follows that either $R'_1 \,/\, \cala_\calh \wbis{\rm b} Q_1 \setminus \cala_\calh$, or
there exist $\bar{Q}_1$ and $Q'_1$ such that $Q_1 \setminus \cala_\calh \warrow{\tau^{*}}{} \bar{Q}_1
\setminus \cala_\calh \arrow{\tau}{} Q'_1 \setminus \cala_\calh$ with $R_1 \, / \, \cala_\calh \wbis{\rm b}
\bar{Q}_1 \setminus \cala_\calh$ and $R'_1 \, / \, \cala_\calh \wbis{\rm b} Q'_1 \setminus \cala_\calh$. In
the former subcase $(Q_1 \pco{L} Q_2) \setminus \cala_\calh$ is allowed to stay idle with $((R'_1 \pco{L}
R_2) / \cala_\calh, (Q_1 \pco{L} Q_2) \setminus \cala_\calh) \in \calb$, while in the latter subcase, since
synchronization does not apply to $\tau$, it follows that $(Q_1 \pco{L} Q_2) \setminus \cala_\calh
\linebreak \warrow{\tau^{*}}{} (\bar{Q}_1 \pco{L} Q_2) \setminus \cala_\calh \arrow{\tau}{} (Q'_1 \pco{L}
Q_2) \setminus \cala_\calh$ with $((R_1 \pco{L} R_2) / \cala_\calh, (\bar{Q}_1 \pco{L} Q_2) \setminus
\cala_\calh) \in \calb$ and $((R'_1 \pco{L} R_2) / \cala_\calh, (Q'_1 \pco{L} Q_2) \setminus \cala_\calh)
\in \calb$.

\item If $(R_1 \pco{L} R_2) \, / \, \cala_\calh \arrow{\tau}{} (R_1 \pco{L} R'_2) \, / \, \cala_\calh$ with
$R_2 \arrow{h}{} R'_2$ and $h \notin L$, then the proof is similar to the one of the previous case.

				\end{itemize}

\item Starting from $(Q \, / \, \cala_\calh) \setminus L$ and $(R \setminus L) \, / \, \cala_\calh$ related
by $\calb$, so that $Q \, / \, \cala_\calh \wbis{\rm b} R \setminus \cala_\calh$, in the branching
bisimulation game there are six cases based on the operational semantic rules in Table~\ref{tab:op_sem}. In
the first three cases, it is $(Q \, / \, \cala_\calh) \setminus L$ to move first:

				\begin{itemize}

\item If $(Q \, / \, \cala_\calh) \setminus L \arrow{l}{} (Q' \, / \, \cala_\calh) \setminus L$ with $Q
\arrow{l}{} Q'$ and $l \notin L$, then $Q \, / \, \cala_\calh \arrow{l}{} Q' \, / \, \cala_\calh$ as $l
\notin \cala_\calh$. From $Q \, / \,\cala_\calh \wbis{\rm b} R \setminus \cala_\calh$ it follows that there
exist $\bar{R}$ and $R'$ such that $R \setminus \cala_\calh \warrow{\tau^{*}}{} \bar{R} \setminus
\cala_\calh \arrow{l}{} R' \setminus \cala_\calh$ with $Q \, / \, \cala_\calh \wbis{\rm b} \bar{R} \setminus
\cala_\calh$ and $Q' \, / \, \cala_\calh \wbis{\rm b} R' \setminus \cala_\calh$. Since neither the
restriction operator nor the hiding operator applies to $\tau$ and $l$, it follows that $(R \setminus L) \,
/ \, \cala_\calh \warrow{\tau^{*}}{} (\bar{R} \setminus L) \, / \, \cala_\calh \arrow{l}{} (R' \setminus L)
\, / \, \cala_\calh$ with $((Q \, / \, \cala_\calh) \setminus L, \linebreak (\bar{R} \setminus L) \, / \,
\cala_\calh) \in \calb$ and $((Q' \, / \, \cala_\calh) \setminus L, (R' \setminus L) \, / \, \cala_\calh)
\in \calb$.

\item If $(Q \, / \, \cala_\calh) \setminus L \arrow{\tau}{} (Q' \, / \, \cala_\calh) \setminus L$ with $Q
\arrow{\tau}{} Q'$, then $Q \, / \, \cala_\calh \arrow{\tau}{} Q' \, / \, \cala_\calh$ as $\tau \notin
\cala_\calh$. \linebreak From $Q \, / \, \cala_\calh \wbis{\rm b} R \setminus \cala_\calh$ it follows that
either $Q' \, / \, \cala_\calh \wbis{\rm b} R \setminus \cala_\calh$, or there exist $\bar{R}$ and $R'$ such
that $R \setminus \cala_\calh \warrow{\tau^{*}}{} \bar{R} \setminus \cala_\calh \arrow{\tau}{} R' \setminus
\cala_\calh$ with $Q \, / \, \cala_\calh \wbis{\rm b} \bar{R} \setminus \cala_\calh$ and $Q' \, / \,
\cala_\calh \wbis{\rm b} R' \setminus \cala_\calh$. In the former subcase $(R \setminus L) \, / \,
\cala_\calh$ is allowed to stay idle with $((Q' \, / \, \cala_\calh) \setminus L, (R \setminus L) \, / \,
\cala_\calh) \in \calb$, while in the latter subcase, since neither the restriction operator nor the hiding
operator applies to $\tau$, it follows that \linebreak $(R \setminus L) \, / \, \cala_\calh
\warrow{\tau^{*}}{} (\bar{R} \setminus L) \, / \, \cala_\calh \arrow{\tau}{} (R' \setminus L) \, / \,
\cala_\calh$ with $((Q \, / \, \cala_\calh) \setminus L, (\bar{R} \setminus L) \, / \, \cala_\calh) \in
\calb$ and $((Q' \, / \, \cala_\calh) \setminus L, (R' \setminus L) \, / \, \cala_\calh) \in \calb$.

\item If $(Q \, / \, \cala_\calh) \setminus L \arrow{\tau}{} (Q' \, / \, \cala_\calh) \setminus L$ with $Q
\arrow{h}{} Q'$, then $Q \, / \, \cala_\calh \arrow{\tau}{} Q' \, / \, \cala_\calh$ as $h \in \cala_\calh$
and the rest of the proof is similar to the one of the previous case.

				\end{itemize}

\noindent
In the other three cases, instead, it is $(R \setminus L) \, / \, \cala_\calh$ to move first:

				\begin{itemize}

\item If $(R \setminus L) \, / \, \cala_\calh \arrow{l}{} (R' \setminus L) \, / \, \cala_\calh$ with $R
\arrow{l}{} R'$ and $l \notin L$, then $R \setminus \cala_\calh \arrow{l}{} R' \setminus \cala_\calh$ as $l
\notin \cala_\calh$. From $R \setminus \cala_\calh \wbis{\rm b}{} Q \, / \, \cala_\calh$ it follows that
there exist $\bar{Q}$ and $Q'$ such that $Q \, / \, \cala_\calh \warrow{\tau^{*}}{} \bar{Q} \, / \,
\cala_\calh \arrow{l}{} Q' \, / \, \cala_\calh$ with $R \setminus \cala_\calh \wbis{\rm b} \bar{Q} \, / \,
\cala_\calh$ and $R' \setminus \cala_\calh \wbis{\rm b} Q' \, / \, \cala_\calh$. Since the restriction
operator does not apply to $\tau$ and $l$, it follows that $(Q \, / \, \cala_\calh) \setminus L \linebreak
\warrow{\tau^{*}}{} (\bar{Q} \, / \, \cala_\calh) \setminus L \arrow{l}{} (Q' \, / \, \cala_\calh) \setminus
L$ with $((R \setminus L) \, / \, \cala_\calh, (\bar{Q} \, / \, \cala_\calh) \setminus L) \in \calb$ and
\linebreak $((R' \setminus L) \, / \, \cala_\calh, (Q' \, / \, \cala_\calh) \setminus L) \in \calb$.

\item If $(R \setminus L) \, / \, \cala_\calh \arrow{\tau}{}(R' \setminus L) \, / \, \cala_\calh$ with $R
\arrow{\tau}{} R'$, then $R \setminus \cala_\calh \arrow{\tau}{} R' \setminus \cala_\calh$ as $\tau \notin
\cala_\calh$. From $R \setminus \cala_\calh \wbis{\rm b}{} Q \, / \, \cala_\calh$ it follows that either $R'
\setminus \cala_\calh \wbis{\rm b}{} Q \, / \, \cala_\calh$, or there exist $\bar{Q}$ and $Q'$ such that $Q
\, / \, \cala_\calh \warrow{\tau^{*}}{} \bar{Q} \, / \, \cala_\calh \arrow{\tau}{} Q' \, / \, \cala_\calh$
with $R \setminus \cala_\calh \wbis{\rm b} \bar{Q} \, / \, \cala_\calh$ and \linebreak $R' \setminus
\cala_\calh \wbis{\rm b} Q' \, / \, \cala_\calh$. In the former subcase $(Q \, / \, \cala_\calh) \setminus
L$ is allowed to stay idle with \linebreak $((R' \setminus L) \, / \, \cala_\calh, (Q \, / \, \cala_\calh)
\setminus L) \in \calb$, while in the latter subcase, since the restriction operator does not apply to
$\tau$, it follows that $(Q \, / \, \cala_\calh) \setminus L \warrow{\tau^{*}}{} (\bar{Q} \, / \,
\cala_\calh) \setminus L \arrow{\tau}{} (Q' \, / \, \cala_\calh) \setminus L$ with $((R \setminus L) \, / \,
\cala_\calh, (\bar{Q} \, / \, \cala_\calh) \setminus L) \in \calb$ and $((R' \setminus L) \, / \,
\cala_\calh, (Q' \, / \, \cala_\calh) \setminus L) \in \calb$.

\item If $(R \setminus L) \, / \, \cala_\calh \arrow{\tau}{} (R' \setminus L) \, / \, \cala_\calh$ with $R
\arrow{h}{} R'$ and $h \notin L$, then $R \, / \, \cala_\calh \arrow{\tau}{} R' \, / \, \cala_\calh$ as $h
\in \cala_\calh$ (note that $R \setminus \cala_\calh$ cannot perform $h$). From $R \, / \, \cala_\calh
\wbis{\rm b} R \setminus \cala_\calh$ -- \linebreak as $P \in \mathrm{SBrSNNI}$ and $R \in \reach(P)$ -- and
$R \setminus \cala_\calh \wbis{\rm b} Q \, / \, \cala_\calh$ it follows that either $R' \, / \, \cala_\calh
\wbis{\rm b} Q \, / \, \cala_\calh$ and hence $R' \setminus \cala_\calh \wbis{\rm b} Q \, / \, \cala_\calh$
-- as $R' \, / \, \cala_\calh \wbis{\rm b} R' \setminus \cala_\calh$ due to $P \in \mathrm{SBrSNNI}$ and $R'
\in \reach(P)$ -- or there exist $\bar{Q}$ and $Q'$ such that $Q \, / \, \cala_\calh \warrow{\tau^{*}}{}
\bar{Q} \, / \, \cala_\calh \arrow{\tau}{} Q' \, / \, \cala_\calh$ with $R \, / \, \cala_\calh \wbis{\rm b}
\bar{Q} \, / \, \cala_\calh$ and $R' \, / \, \cala_\calh \wbis{\rm b} Q' \, / \, \cala_\calh$ and hence $R
\setminus \cala_\calh \wbis{\rm b} \bar{Q} \, / \, \cala_\calh$ and $R' \setminus \cala_\calh \wbis{\rm b}
Q' \, / \, \cala_\calh$. In the former subcase $(Q \, / \, \cala_\calh) \setminus L$ is allowed to stay idle
with $((R' \setminus L) \, / \, \cala_\calh, (Q \, / \, \cala_\calh) \setminus L) \in \calb$, while in the
latter subcase, since the restriction operator does not apply to $\tau$, it follows that $(Q \, / \,
\cala_\calh) \setminus L \warrow{\tau^{*}}{} (\bar{Q} \, / \, \cala_\calh) \setminus L \arrow{\tau}{} (Q' \,
/ \, \cala_\calh) \setminus L$ with $((R \setminus L) \, / \, \cala_\calh , (\bar{Q} \, / \, \cala_\calh)
\setminus L) \in \calb$ and $((R' \setminus L) \, / \, \cala_\calh , (Q' \, / \, \cala_\calh) \setminus L)
\in \calb$.

				\end{itemize}

\item Starting from $((Q_1 \pco{L} Q_2) \setminus \cala_\calh, (R_1 \pco{L} R_2) \setminus \cala_\calh) \in
\calb$, so that $Q_1 \setminus \cala_\calh \wbis{\rm b} R_1 \setminus \cala_\calh$ and $Q_2 \setminus
\cala_\calh \wbis{\rm b} R_2 \setminus \cala_\calh$, in the branching bisimulation game there are five cases
based on the operational semantic rules in Table~\ref{tab:op_sem}:

				\begin{itemize}

\item If $(Q_1 \pco{L} Q_2) \setminus \cala_\calh \arrow{l}{} (Q'_1 \pco{L} Q_2) \setminus \cala_\calh$ with
$Q_1 \arrow{l}{} Q'_1$ and $l \notin L$, then $Q_1 \setminus \cala_\calh \linebreak \arrow{l}{} Q'_1
\setminus \cala_\calh$ as $l \notin \cala_\calh$. From $Q_1 \setminus \cala_\calh \wbis{\rm b} R_1 \setminus
\cala_\calh$ it follows that there exist $\bar{R}_1$ and $R'_1$ such that $R_1 \setminus \cala_\calh
\warrow{\tau^{*}}{} \bar{R}_1 \setminus \cala_\calh \arrow{l}{} R'_1 \setminus \cala_\calh$ with $Q_1
\setminus \cala_\calh \wbis{\rm b} \bar{R}_1 \setminus \cala_\calh$ and $Q'_1 \setminus \cala_\calh
\wbis{\rm b} R'_1 \setminus \cala_\calh$. Since synchronization does not apply to $\tau$, it follows that
$(R_1 \pco{L} R_2) \setminus \cala_\calh \warrow{\tau^{*}}{} (\bar{R}_1 \pco{L} R_2) \setminus \cala_\calh
\arrow{l}{} (R'_1 \pco{L} R_2) \setminus \cala_\calh$ with $((Q_1 \pco{L} Q_2) \setminus \cala_\calh,
\linebreak (\bar{R}_1 \pco{L} R_2) \setminus \cala_\calh) \in \calb$ and $((Q'_1 \pco{L} Q_2) \setminus
\cala_\calh, (R'_1 \pco{L} R_2) \setminus \cala_\calh) \in \calb$.

\item If $(Q_1 \pco{L} Q_2) \setminus \cala_\calh \arrow{l}{} (Q_1 \pco{L} Q'_2) \setminus \cala_\calh$ with
$Q_2 \arrow{l}{} Q'_2$ and $l \notin L$, then the proof is similar to the one of the previous case.

\item If $(Q_1 \pco{L} Q_2) \setminus \cala_\calh \arrow{l}{} (Q'_1 \pco{L} Q'_2) \setminus \cala_\calh$
with $Q_i \arrow{l}{} Q'_i$ for $i \in \{ 1, 2 \}$ and $l \in L$, then $Q_i \setminus \cala_\calh
\arrow{l}{} Q'_i \setminus \cala_\calh$ as $l \notin \cala_\calh$. From $Q_i \setminus \cala_\calh \wbis{\rm
b} R_i \setminus \cala_\calh$ it follows that there exist $\bar{R}_i$ and $R'_i$ such that $R_i \setminus
\cala_\calh \warrow{\tau^{*}}{} \bar{R}_i \setminus \cala_\calh \arrow{l}{} R'_i \setminus \cala_\calh$ with
$Q_i \setminus \cala_\calh \wbis{\rm b} \bar{R}_i \setminus \cala_\calh$ and $Q'_i \setminus \cala_\calh
\wbis{\rm b} R'_i \setminus \cala_{\calh}$. Since synchronization does not apply to $\tau$, it follows that
$(R_1 \pco{L} R_2) \setminus \cala_\calh \warrow{\tau^{*}}{} (\bar{R}_1 \pco{L} \bar{R}_2) \setminus
\cala_\calh \arrow{l}{} (R'_1 \pco{L} R'_2) \setminus \cala_\calh$ with $((Q_1 \pco{L} Q_2) \setminus
\cala_\calh, \linebreak (\bar{R}_1 \pco{L} \bar{R}_2) \setminus \cala_\calh) \in \calb$ and $((Q'_1 \pco{L}
Q'_2) \setminus \cala_\calh, (R'_1 \pco{L} R'_2) \setminus \cala_\calh) \in \calb$.

\item If $(Q_1 \pco{L} Q_2) \setminus \cala_\calh \arrow{\tau}{} (Q'_1 \pco{L} Q_2) \setminus \cala_\calh$
with $Q_1 \arrow{\tau}{} Q'_1$, then $Q_1 \setminus \cala_\calh \arrow{\tau}{} Q'_1 \setminus \cala_\calh$
as $\tau \notin \cala_\calh$. From $Q_1 \setminus \cala_\calh \wbis{\rm b} R_1 \setminus\cala_\calh$ it
follows that either $Q'_1 \setminus \cala_\calh \wbis{\rm b} R_1 \setminus \cala_\calh$, or there exist
$\bar{R}_1$ and $R'_1$ such that $R_1 \setminus \cala_\calh \warrow{\tau^{*}}{} \bar{R}_1 \setminus
\cala_\calh \arrow{\tau}{} R'_1 \setminus \cala_\calh$ with $Q_1 \setminus \cala_\calh \wbis{\rm b}
\bar{R}_1 \setminus \cala_\calh$ and $Q'_1 \setminus \cala_\calh \wbis{\rm b} R'_1 \setminus \cala_\calh$.
In the former subcase $(R_1 \pco{L} R_2) \setminus \cala_\calh$ is allowed to stay idle with $((Q'_1 \pco{L}
Q_2) \setminus \cala_\calh, (R_1 \pco{L} R_2) \setminus \cala_\calh) \in \calb$, while in the latter
subcase, since synchronization does not apply to $\tau$, it follows that $(R_1 \pco{L} R_2) \setminus
\cala_\calh \warrow{\tau^{*}}{} (\bar{R}_1 \pco{L} R_2) \setminus \cala_\calh \linebreak \arrow{\tau}{}
(R'_1 \pco{L} R_2) \setminus \cala_\calh$ with $((Q_1 \pco{L} Q_2) \setminus \cala_\calh, (\bar{R}_1 \pco{L}
R_2) \setminus \cala_\calh) \in \calb$ and $((Q'_1 \pco{L} Q_2) \setminus \cala_\calh, \linebreak (R'_1
\pco{L} R_2) \setminus \cala_\calh) \in \calb$.

\item If $(Q_1 \pco{L} Q_2) \setminus \cala_\calh \arrow{\tau}{} (Q_1 \pco{L} Q'_2) \setminus \cala_\calh$
with $Q_2 \arrow{\tau}{} Q'_2$, then the proof is similar to the one of the previous case.
\qedhere

				\end{itemize}

			\end{enumerate}

		\end{proof}

	\end{lem}

	\begin{thm}\label{thm:compositionality} 

Let $P, P_1, P_2 \in \procs$ and $\calp \in \{ \mathrm{SBrSNNI}, \mathrm{P\_BrNDC}, \mathrm{SBrNDC} \}$.
Then:

		\begin{enumerate}

\item $P \in \calp \Longrightarrow a \, . \, P \in \calp$ for all $a \in \cala_\call \cup \{ \tau \}$.

\item $P_1, P_2 \in \calp \Longrightarrow P_1 \pco{L} P_2 \in \calp$ for all $L \subseteq \cala_\call$ if
$\calp \in \{ \mathrm{SBrSNNI}, \mathrm{P\_BrNDC} \}$ or for all $L \subseteq \cala$ if $\calp =
\mathrm{SBrNDC}$.

\item $P \in \calp \Longrightarrow P \setminus L \in \calp$ for all $L \subseteq \cala$.

\item $P \in \calp \Longrightarrow P \, / \, L \in \calp$ for all $L \subseteq \cala_\call$.
\fullbox

		\end{enumerate}

		\begin{proof}
We first prove the four results for SBrSNNI, from which it will follow that they hold for P\_BrNDC too by
virtue of the forthcoming Theorem~\ref{thm:branching_taxonomy_1}:

			\begin{enumerate}

\item Given an arbitrary $P \in \mathrm{SBrSNNI}$ and an arbitrary $a \in \cala_\call \cup \{ \tau \}$, from
$P \setminus \cala_\calh \wbis{\rm b} P \, / \, \cala_\calh$ we derive that $a \, . \, (P \setminus
\cala_\calh) \wbis{\rm b} a \, . \, (P \, / \, \cala_\calh)$ because $\wbis{\rm b}$ is a congruence with
respect to action prefix~\cite{GW96}, from which it follows that $(a \, . \, P) \setminus \cala_\calh
\wbis{\rm b} (a \, . \, P) \, / \, \cala_\calh$, i.e., $a \, . \, P \in \mathrm{BrSNNI}$, because $a \notin
\cala_\calh$. To conclude the proof, it suffices to observe that all the processes reachable from $a \, . \,
P$ after performing $a$ are processes reachable from $P$, which are known to be BrSNNI.

\item Given two arbitrary $P_1, P_2 \in \mathrm{SBrSNNI}$ and an arbitrary $L \subseteq \cala_\call$, the
result follows from Lemma~\ref{lem:compositionality}(1) by taking $Q_1$ identical to $R_1$ and $Q_2$
identical to $R_2$.

\item Given an arbitrary $P \in \mathrm{SBrSNNI}$ and an arbitrary $L \subseteq \cala$, the result follows
from Lemma~\ref{lem:compositionality}(2) by taking $Q$ identical to $R$ -- which will be denoted by $P'$ --
because:

				\begin{itemize}

\item $(P' \setminus L) \setminus \cala_\calh \wbis{\rm b} (P' \setminus \cala_\calh) \setminus L$ as the
order in which restriction sets are considered \linebreak is unimportant.

\item $(P' \setminus \cala_\calh) \setminus L \wbis{\rm b} (P' \, / \, \cala_\calh) \setminus L$ due to $P'
\setminus \cala_\calh \wbis{\rm b} P' \, / \, \cala_\calh$ -- as $P \in \mathrm{SBrSNNI}$ and $P' \in
\reach(P)$ -- and $\wbis{\rm b}$ being a congruence with respect to the restriction operator due to
Lemma~\ref{lem:congr_restr_hiding_parallel}.

\item $(P' \, / \, \cala_\calh) \setminus L \wbis{\rm b} (P' \setminus L) \, / \, \cala_\calh$ as shown in
Lemma~\ref{lem:compositionality}(2).

\item From the transitivity of $\wbis{\rm b}$ we obtain that $(P' \setminus L) \setminus \cala_\calh
\wbis{\rm b} (P' \setminus L) \, / \, \cala_\calh$.

				\end{itemize}

\item Given an arbitrary $P \in \mathrm{SBrSNNI}$ and an arbitrary $L \subseteq \cala_\call$, for every $P'
\in \reach(P)$ \linebreak it holds that $P' \setminus \cala_\calh \wbis{\rm b} P' \, / \, \cala_\calh$, from
which we derive that $(P' \setminus \cala_\calh) \, / \, L \wbis{\rm b} (P' / \cala_\calh) \, / \, L$
because $\wbis{\rm b}$ is a congruence with respect to the hiding operator due to
Lemma~\ref{lem:congr_restr_hiding_parallel}. Since $L \cap \cala_{\calh} = \emptyset$, we have that $(P'
\setminus \cala_\calh) \, / \, L$ is isomorphic to $(P' \, / \, L) \setminus \cala_\calh$ and $(P' \, / \,
\cala_\calh) \, / \, L$ is isomorphic to $(P' \, / \, L) \, / \, \cala_\calh$, hence $(P' \, / \, L)
\setminus \cala_\calh \wbis{\rm b} (P' \, / \, L) \, / \, \cala_\calh$, i.e., $P' \, / \, L$ is BrSNNI.

			\end{enumerate}

\noindent
We then prove the four results for SBrNDC:

			\begin{enumerate}

\item Given an arbitrary $P \in \mathrm{SBrNDC}$ and an arbitrary $a \in \cala_\tau \setminus \cala_\calh$,
it trivially holds that $a \, . \, P \in \mathrm{SBrNDC}$ because $a$ is not high and all the processes
reachable from $a \, . \, P$ after performing $a$ are processes reachable from $P$, which is known to be
SBrNDC.

\item Given two arbitrary $P_1, P_2 \in \mathrm{SBrNDC}$ and an arbitrary $L \subseteq \cala$, the result
follows from Lemma~\ref{lem:compositionality}(3) as can be seen by observing that whenever $P'_1 \pco{L}
P'_2 \arrow{h}{} P''_1 \pco{L} P''_2$ for $P'_1 \pco{L} P'_2 \in \reach(P_1 \pco{L} P_2)$:

				\begin{itemize}

\item If $P'_1 \arrow{h}{} P''_1$, $P''_2 = P'_2$, and $h \notin L$, then from $P_1 \in \mathrm{SBrNDC}$ it
follows that $P'_1 \setminus \cala_\calh \wbis{\rm b} P''_1 \setminus \cala_\calh$ and hence $(P'_1 \pco{L}
P'_2) \setminus \cala_\calh \wbis{\rm b} (P''_1 \pco{L} P''_2) \setminus \cala_{\calh}$ as $P'_2 \setminus
\cala_{\calh} \wbis{\rm b} P''_2 \setminus \cala_{\calh}$.

\item If $P'_2 \arrow{h}{} P''_2$, $P''_1 = P'_1$, and $h \notin L$, then from $P_2 \in \mathrm{SBrNDC}$ it
follows that $P'_2 \setminus \cala_\calh \wbis{\rm b} P''_2 \setminus \cala_\calh$ and hence $(P'_1 \pco{L}
P'_2) \setminus \cala_\calh \wbis{\rm b} (P''_1 \pco{L} P''_2) \setminus \cala_{\calh}$ as $P'_1 \setminus
\cala_{\calh} \wbis{\rm b} P''_1 \setminus \cala_{\calh}$.

\item If $P'_1 \arrow{h}{} P''_1$, $P'_2 \arrow{h}{} P''_2$, and $h \in L$, then from $P_1, P_2 \in
\mathrm{SBrNDC}$ it follows that $P'_1 \setminus \cala_\calh \wbis{\rm b} P''_1 \setminus \cala_\calh$ and
$P'_2 \setminus \cala_\calh \wbis{\rm b} P''_2 \setminus \cala_\calh$, which in turn entail that $(P'_1
\pco{L} P'_2) \setminus \cala_\calh \linebreak \wbis{\rm b} (P''_1 \pco{L} P''_2) \setminus \cala_\calh$.

				\end{itemize}

\item Given an arbitrary $P \in \mathrm{SBrNDC}$ and an arbitrary $L \subseteq \cala$, for every $P' \in
\reach(P)$ and for every $P''$ such that $P' \arrow{h}{} P''$ it holds that $P' \setminus \cala_\calh
\wbis{\rm b} P'' \setminus \cala_\calh$, from which we derive that $(P' \setminus \cala_\calh) \setminus L
\wbis{\rm b} (P'' \setminus \cala_\calh) \setminus L$ because $\wbis{\rm b}$ is a congruence with respect to
the restriction operator due to Lemma~\ref{lem:congr_restr_hiding_parallel}. Since $(P' \setminus
\cala_\calh) \setminus L$ is isomorphic to $(P' \setminus L) \setminus \cala_\calh$ and $(P'' \setminus
\cala_\calh) \setminus L$ is isomorphic to $(P'' \setminus L) \setminus \cala_\calh$, we have that $(P'
\setminus L) \setminus \cala_\calh \wbis{\rm b} (P'' \setminus L) \setminus \cala_\calh$.

\item Given an arbitrary $P \in \mathrm{SBrNDC}$ and an arbitrary $L \subseteq \cala_\call$, for every $P'
\in \reach(P)$ and for every $P''$ such that $P' \arrow{h}{} P''$ it holds that $P' \setminus \cala_\calh
\wbis{\rm b} P'' \setminus \cala_\calh$, from which we derive that $(P' \setminus \cala_\calh) \, / \, L
\wbis{\rm b} (P'' \setminus \cala_\calh) \, / \, L$ because $\wbis{\rm b}$ is a congruence with respect to
the hiding operator due to Lemma~\ref{lem:congr_restr_hiding_parallel}. Since $L \cap \cala_{\calh} =
\emptyset$, we have that $(P' \setminus \cala_\calh) \, / \, L$ is isomorphic to $(P' \, / \, L) \setminus
\cala_\calh$ and $(P'' \setminus \cala_\calh) \, / \, L$ is isomorphic to $(P'' \, / \, L) \setminus
\cala_\calh$, hence $(P' \, / \, L) \setminus \cala_\calh \wbis{\rm b} (P'' \, / \, L) \setminus
\cala_\calh$.
\qedhere

			\end{enumerate}

		\end{proof}

	\end{thm}

%
\subsection{Taxonomy of Security Properties}
\label{sec:branching_taxonomy} 
%

The relationships among the various $\wbis{\rm b}$-based noninterference properties turn out to follow the
same pattern as Theorem~\ref{thm:weak_bisim_taxonomy}.

In~\cite{EAB23} some parts of the proof of the forthcoming Theorem~\ref{thm:branching_taxonomy_1} -- as well
as some parts of the proof of the previous Theorem~\ref{thm:compositionality} -- proceeded by induction on
the depth of the labeled transition system underlying the process under examination. Now that the language
includes recursion, which may introduce cycles, we have to follow a different proof technique, which relies
on the notion of branching bisimulation up to $\wbis{\rm b}$ of~\cite{Gla93} recalled below.

	\begin{defi}\label{def:up_to}

A symmetric binary relation $\calb$ over $\procs$ is a \emph{branching bisimulation up to~$\wbis{\rm b}$}
iff, whenever $(P_{1}, P_{2}) \in \calb$, then:

		\begin{itemize}

\item for each $P_{1} \warrow{\tau^{*}}{} \bar{P}_{1} \arrow{a}{} P'_{1}$ with $P_{1} \wbis{\rm b}
\bar{P}_{1}$:

			\begin{itemize}

\item either $a = \tau$ and $\bar{P}_{1} \wbis{\rm b} P'_{1}$;

\item or there exists $P_{2} \warrow{\tau^{*}}{} \bar{P}_{2} \arrow{a}{} P'_{2}$ such that $\bar{P}_{1}
\wbis{\rm b} \calb \wbis{\rm b} \bar{P}_{2}$ and $P'_{1} \wbis{\rm b} \calb \wbis{\rm b} P'_{2}$.
\fullbox

			\end{itemize}

		\end{itemize}

	\end{defi}

\noindent
In the definition above, $\wbis{\rm b} \calb \wbis{\rm b}$ stands for the composition of the three mentioned
relations. Moreover, in the case that $a = \tau$ and $\bar{P}_{1} \wbis{\rm b} P'_{1}$, since the considered
relations are symmetric and $\wbis{\rm b}$ is also transitive and reflexive, it holds that $P'_{1} \wbis{\rm
b} \bar{P}_{1} \wbis{\rm b} P_{1} \; \calb \; P_{2} \wbis{\rm b} P_{2}$, i.e., $P'_{1} \wbis{\rm b} \calb
\wbis{\rm b} P_{2}$. As shown in~\cite{Gla93}, if $\calb$ is a branching bisimulation up to $\wbis{\rm b}$
and $(P_{1}, P_{2}) \in \calb$, then $P_{1} \wbis{\rm b} P_{2}$ because $\wbis{\rm b} \calb \wbis{\rm b}$
turns out to be a branching bisimulation; \linebreak the advantage of working with $\calb$ is that it needs
to include fewer pairs. While in~\cite{FG01} weak bisimulation up to $\wbis{}$~\cite{SM92} has been
exploited several times to prove various results, here branching bisimulation up to $\wbis{\rm b}$ is
employed only to show that $\mathrm{SBrNDC} \subset \mathrm{SBrSNNI}$.

To study the taxonomy of the noninterference properties in Definition~\ref{def:branching_bisim_properties},
we first prove some further ancillary results about parallel composition, restriction, and hiding under
SBrSNNI and SBrNDC.

	\begin{lem}\label{lem:branching_taxonomy}

Let $P, P_1, P_2 \in \procs$. Then:

		\begin{enumerate}

\item If $P \in \mathrm{SBrNDC}$ and $P' \, / \, \cala_\calh \warrow{\tau^{*}}{} P'' \, / \, \cala_\calh$
for $P' \in \reach(P)$, then $P' \setminus \cala_\calh \warrow{\tau^{*}}{} \hat{P}'' \setminus \cala_\calh$
with $P'' \setminus \cala_\calh \wbis{\rm b} \hat{P}'' \setminus \cala_\calh$.

\item If $P_1, P_2 \in \mathrm{SBrNDC}$ and $P_1 \setminus \cala_\calh \wbis{\rm b} P_2 \setminus
\cala_\calh$, then $P_1 \, / \, \cala_\calh \wbis{\rm b} P_2 \, / \, \cala_\calh$.

\item If $P_2 \in \mathrm{SBrSNNI}$ and $L \subseteq \cala_\calh$, then $P'_1 \setminus \cala_\calh
\wbis{\rm b} ((P'_2 \pco{L} Q) \, / \, L) \setminus \cala_\calh$ for all $Q \in \procs$ having only actions
in $\cala_\calh$ and for all $P'_1 \in \reach(P_1)$ and $P'_2 \in \reach(P_2)$ such that $P'_1 \setminus
\cala_\calh \wbis{\rm b} P'_2 \, / \, \cala_\calh$.

		\end{enumerate}

		\begin{proof}
Let $\calb$ be a symmetric relation containing all the pairs of processes that have to be shown to be
branching bisimilar according to the property considered between the last two stated above:

			\begin{enumerate}

\item We proceed by induction on the number $n \in \natns$ of $\tau$-transitions in $P' \, / \, \cala_\calh
\warrow{\tau^{*}}{} P'' \, / \, \cala_\calh$:

				\begin{itemize}

\item If $n = 0$ then $P'\, / \, \cala_\calh$ stays idle and $P''\, / \, \cala_\calh$ is $P'\, / \,
\cala_\calh$. Likewise, $P' \setminus \cala_\calh$ can stay idle, i.e., $P' \setminus \cala_\calh
\warrow{\tau^{*}}{} P' \setminus \cala_\calh$, with $P' \setminus \cala_\calh \wbis{\rm b} P' \setminus
\cala_\calh$ as $\wbis{\rm b}$ is reflexive.

\item Let $n > 0$ and $P'_0 \, / \, \cala_\calh \arrow{\tau}{} P'_1 \, / \, \cala_\calh \arrow{\tau}{} \dots
\arrow{\tau}{} P'_{n - 1} \, / \, \cala_\calh \arrow{\tau}{} P'_n \, / \, \cala_\calh$ where $P'_0 \, / \,
\cala_\calh$ is $P' \, / \, \cala_\calh$ and $P'_n \, / \, \cala_\calh$ is $P'' \, / \, \cala_\calh$. From
the induction hypothesis it follows that $P' \setminus \cala_\calh \warrow{\tau^{*}}{} \hat{P}'_{n - 1}
\setminus \cala_\calh$ with $P'_{n - 1} \setminus \cala_\calh \wbis{\rm b} \hat{P}'_{n - 1} \setminus
\cala_\calh$. As far as the $n$-th $\tau$-transition $P'_{n - 1} \, / \, \cala_\calh \arrow{\tau}{} P'_n \,
/ \, \cala_\calh$ is concerned, there are two cases depending on whether it is originated from $P'_{n - 1}
\arrow{\tau}{} P'_n$ or $P'_{n - 1} \arrow{h}{} P'_n$:

					\begin{itemize}

\item If $P'_{n - 1} \arrow{\tau}{} P'_n$ then $P'_{n - 1} \setminus \cala_\calh \arrow{\tau}{} P'_n
\setminus \cala_\calh$. Since $P'_{n - 1} \setminus \cala_\calh \wbis{\rm b} \hat{P}'_{n - 1} \setminus
\cala_\calh$, \linebreak it follows that:

						\begin{itemize}

\item Either $P'_{n} \setminus \cala_\calh \wbis{\rm b} \hat{P}'_{n - 1} \setminus \cala_\calh$, in which
case $\hat{P}'_{n - 1} \setminus \cala_\calh$ stays idle and we are done because $P' \setminus \cala_\calh
\warrow{\tau^{*}}{} \hat{P}'_{n - 1} \setminus \cala_\calh$.

\item Or $\hat{P}'_{n - 1} \setminus \cala_\calh \warrow{\tau^{*}}{} \hat{P}''_{n - 1} \setminus \cala_\calh
\arrow{\tau}{} \hat{P}'_{n} \setminus \cala_\calh$ with $P'_{n - 1} \setminus \cala_\calh \wbis{\rm b}
\hat{P}''_{n - 1} \setminus \cala_\calh$ and $P'_{n} \setminus \cala_\calh \wbis{\rm b} \hat{P}'_{n}
\setminus \cala_\calh$, in which case we are done because $P' \setminus \cala_\calh \warrow{\tau^{*}}{}
\hat{P}'_n \setminus \cala_\calh$.

						\end{itemize}

\item If $P'_{n - 1} \arrow{h}{} P'_n$ then from $P \in \mathrm{SBrNDC}$ it follows that $P'_{n - 1}
\setminus \cala_\calh \wbis{\rm b} P'_{n} \setminus \cala_\calh$. Since $P'_{n - 1} \setminus \cala_\calh
\wbis{\rm b} \hat{P}'_{n - 1} \setminus \cala_\calh$ and $\wbis{\rm b}$ is symmetric and transitive, we
obtain $P'_{n} \setminus \cala_\calh \wbis{\rm b} \hat{P}'_{n - 1} \setminus \cala_\calh$. Thus we are done
because $P' \setminus \cala_\calh \warrow{\tau^{*}}{} \hat{P}'_{n - 1} \setminus \cala_\calh$.

					\end{itemize}

				\end{itemize}

\item Starting from $(P_1 \, / \, \cala_\calh, P_2 \, / \, \cala_\calh) \in \calb$, so that $P_1 \setminus
\cala_\calh \wbis{\rm b} P_2 \setminus \cala_\calh$, in the branching bisimulation game there are three
cases based on the operational semantic rules in Table~\ref{tab:op_sem}:

				\begin{itemize}

\item If $P_1 \, / \, \cala_\calh \arrow{\tau}{} P'_1 \, / \, \cala_\calh$ with $P_1 \arrow{h}{} P'_1$, then
$P_1 \setminus \cala_\calh \wbis{\rm b} P'_1 \setminus \cala_\calh$ as $h \in \cala_\calh$ and $P_1 \in
\mathrm{SBrNDC}$. Since $P_1 \setminus \cala_\calh \wbis{\rm b} P_2 \setminus \cala_\calh$, so that $P'_1
\setminus \cala_\calh \wbis{\rm b} P_2 \setminus \cala_\calh$ as $\wbis{\rm b}$ is symmetric and transitive,
and $P'_1, P_2 \in \mathrm{SBrNDC}$, it follows that $P_2 \, / \, \cala_\calh$ is allowed to stay idle with
$(P'_1 \, / \, \cala_\calh, P_2 \, / \, \cala_\calh) \in \calb$.

\item If $P_1 \, / \, \cala_\calh \arrow{l}{} P'_1 \, / \, \cala_\calh$ with $P_1 \arrow{l}{} P'_1$, then
$P_1 \setminus \cala_\calh \arrow{l}{} P'_1 \setminus \cala_\calh$ as $l \notin \cala_\calh$. From $P_1
\setminus \cala_\calh \wbis{\rm b} P_2 \setminus \cala_\calh$ it follows that there exist $\bar{P}_2$ and
$P'_2$ such that $P_2 \setminus \cala_\calh \warrow{\tau^{*}}{} \bar{P}_2 \setminus \cala_\calh \linebreak
\arrow{l}{} P'_2 \setminus \cala_\calh$ with $P_1 \setminus \cala_\calh \wbis{\rm b} \bar{P}_2 \setminus
\cala_\calh$ and $P'_1 \setminus \cala_\calh \wbis{\rm b} P'_2 \setminus \cala_{\calh}$. Thus $P_2 \, /
\cala_\calh \warrow{\tau^{*}}{} \bar{P}_2 \, / \cala_\calh \linebreak \arrow{l}{} P'_2 \, / \, \cala_\calh$.
Since $P_1 \setminus \cala_\calh \wbis{\rm b} \bar{P}_2 \setminus \cala_\calh$ with $P_1, \bar{P}_2 \in
\mathrm{SBrNDC}$ and $P'_1 \setminus \cala_\calh \wbis{\rm b} P'_2 \setminus \cala_{\calh}$ with $P'_1, P'_2
\in \mathrm{SBrNDC}$, we have $(P_1 \, / \, \cala_\calh, \bar{P}_2 \, / \, \cala_\calh) \in \calb$ and
$(P'_1 \, / \, \cala_\calh, P'_2 \, / \, \cala_\calh) \in \calb$.

\item If $P_1 \, / \, \cala_\calh \arrow{\tau}{} P'_1 \, / \, \cala_\calh$ with $P_1 \arrow{\tau}{} P'_1$,
then the proof is similar to the previous one, with the additional possibility that, in response to $P_1
\setminus \cala_\calh \arrow{\tau}{} P'_1 \setminus \cala_\calh$, $P_2 \setminus \cala_\calh$ stays idle
with $P'_1 \setminus \cala_\calh \wbis{\rm b} P_2 \setminus \cala_{\calh}$, so that $P_2 \, / \,
\cala_\calh$ stays idle too with $(P'_1 \, / \, \cala_\calh, P_2 \, / \, \cala_\calh) \in \calb$ because
$P'_1 \setminus \cala_\calh \wbis{\rm b} P_2 \setminus \cala_{\calh}$ and $P'_1, P_2 \in \mathrm{SBrNDC}$.

				\end{itemize}

\item Starting from $P'_1 \setminus \cala_\calh$ and $((P'_2 \pco{L} Q) \, / \, L) \setminus \cala_\calh$
related by $\calb$, so that $P'_1 \setminus \cala_\calh \wbis{\rm b} P'_2 \, / \, \cala_\calh$, in the
branching bisimulation game there are six cases based on the operational semantic rules in
Table~\ref{tab:op_sem}. In the first two cases, it is $P'_1 \setminus \cala_\calh$ to move first:

				\begin{itemize}

\item If $P'_1 \setminus \cala_\calh \arrow{l}{} P''_1 \setminus \cala_\calh$ we observe that from $P'_2 \in
\reach(P_2)$ and $P_2 \in \mathrm{SBrSNNI}$ it follows that $P'_2 \setminus \cala_\calh \wbis{\rm b} P'_2 \,
/ \, \cala_\calh$, so that $P'_1 \setminus \cala_\calh \wbis{\rm b} P'_2 \, / \, \cala_\calh \wbis{\rm b}
P'_2 \setminus \cala_\calh$, i.e., $P'_1 \setminus \cala_\calh \wbis{\rm b} P'_2 \setminus \cala_\calh$, as
$\wbis{\rm b}$ is symmetric and transitive. As a consequence, since $l \neq \tau$ there exist $\bar{P}'_2$
and $P''_2$ such that $P'_2 \setminus \cala_\calh \warrow{\tau^{*}}{} \bar{P}'_2 \setminus \cala_\calh
\arrow{l}{} P''_2 \setminus \cala_\calh$ with $P'_1 \setminus \cala_\calh \wbis{\rm b} \bar{P}'_2 \setminus
\cala_\calh$ and $P''_1 \setminus \cala_\calh \wbis{\rm b} P''_2 \setminus \cala_\calh$. Thus, $((P'_2
\pco{L} Q) \, / \, L) \setminus \cala_\calh \warrow{\tau^{*}}{} ((\bar{P}'_2 \pco{L} Q) \, / \, L) \setminus
\cala_\calh \linebreak \arrow{l}{} ((P''_2 \pco{L} Q) \, / \, L) \setminus \cala_\calh$ with $(P'_1
\setminus \cala_\calh, ((\bar{P}'_2 \pco{L} Q) \, / \, L) \setminus \cala_\calh) \in \calb$ -- because
\linebreak $P'_1 \in \reach(P_1)$, $\bar{P}'_2 \in \reach(P_2)$, and $P'_1 \setminus \cala_\calh \wbis{\rm
b} \bar{P}'_2 \, / \, \cala_\calh$ as $P_2 \in \mathrm{SBrSNNI}$ -- and $(P''_1 \setminus \cala_\calh,
((P''_2 \pco{L} Q) \, / \, L) \setminus \cala_\calh) \in \calb$ -- because $P''_1 \in \reach(P_1)$, $P''_2
\in \reach(P_2)$, and $P''_1 \setminus \cala_\calh \wbis{\rm b} P''_2 \, / \, \cala_\calh$ as $P_2 \in
\mathrm{SBrSNNI}$.

\item If $P'_1 \setminus \cala_\calh \arrow{\tau}{} P''_1 \setminus \cala_\calh$ there are two subcases:

					\begin{itemize}

\item If $P''_1 \setminus \cala_\calh \wbis{\rm b} P'_2 \, / \, \cala_\calh$ then $(P'_2 \pco{L} Q) \, / \,
L) \setminus \cala_\calh$ is allowed to stay idle with \linebreak $(P''_1 \setminus \cala_\calh, ((P'_2
\pco{L} Q) \, / \, L) \setminus \cala_\calh) \in \calb$ because $P''_1 \in \reach(P_1)$ and $P'_2 \in
\reach(P_2)$.

\item If $P''_1 \setminus \cala_\calh \not\wbis{\rm b} P'_2 \, / \, \cala_\calh$ we observe that from $P'_2
\in \reach(P_2)$ and $P_2 \in \mathrm{SBrSNNI}$ \linebreak it follows that $P'_2 \setminus \cala_\calh
\wbis{\rm b} P'_2 \, / \, \cala_\calh$, so that on the one hand $P'_1 \setminus \cala_\calh \wbis{\rm b}
P'_2 \, / \, \cala_\calh \wbis{\rm b} P'_2 \setminus \cala_\calh$, i.e., $P'_1 \setminus \cala_\calh
\wbis{\rm b} P'_2 \setminus \cala_\calh$, while on the other hand $P''_1 \setminus \cala_\calh \not\wbis{\rm
b} P'_2 \, / \, \cala_\calh \wbis{\rm b} P'_2 \setminus \cala_\calh$, i.e., $P''_1 \setminus \cala_\calh
\not\wbis{\rm b} P'_2 \setminus \cala_\calh$. As a consequence, there exist $\bar{P}'_2$ and $P''_2$ such
that $P'_2 \setminus \cala_\calh \warrow{\tau^{*}}{} \bar{P}'_2 \setminus \cala_\calh \arrow{\tau}{} P''_2
\setminus \cala_\calh$ with $P'_1 \setminus \cala_\calh \wbis{\rm b} \bar{P}'_2 \setminus \cala_\calh$ and
$P''_1 \setminus \cala_\calh \wbis{\rm b} P''_2 \setminus \cala_\calh$. Therefore, $((P'_2 \pco{L} Q) \, /
\, L) \setminus \cala_\calh \warrow{\tau^{*}}{} ((\bar{P}'_2 \pco{L} Q) \, / \, L) \setminus \cala_\calh
\arrow{\tau}{} ((P''_2 \pco{L} Q) \, / \, L) \setminus \cala_\calh$ with $(P'_1 \setminus \cala_\calh,
((\bar{P}'_2 \pco{L} Q) \, / \, L) \setminus \cala_\calh) \in \calb$ -- because $P'_1 \in \reach(P_1)$,
$\bar{P}'_2 \in \reach(P_2)$, and $P'_1 \setminus \cala_\calh \wbis{\rm b} \bar{P}'_2 \, / \, \cala_\calh$
as $P_2 \in \mathrm{SBrSNNI}$ -- and $(P''_1 \setminus \cala_\calh, ((P''_2 \pco{L} Q) \, / \, L) \setminus
\cala_\calh) \in \calb$ \linebreak -- because $P''_1 \in \reach(P_1)$, $P''_2 \in \reach(P_2)$, and $P''_1 \setminus
\cala_\calh \wbis{\rm b} P''_2 \, / \, \cala_\calh$ \linebreak as $P_2 \in \mathrm{SBrSNNI}$.

					\end{itemize}

				\end{itemize}

\noindent
In the other four cases, instead, it is $((P'_2 \pco{L} Q) \, / \, L) \setminus \cala_\calh$ to move first:

				\begin{itemize}

\item If $((P'_2 \pco{L} Q) \, / \, L) \setminus \cala_\calh \arrow{l}{} ((P''_2 \pco{L} Q) \, / \, L)
\setminus \cala_\calh$ with $P'_2 \arrow{l}{} P''_2$ so that $P'_2 \setminus \cala_\calh \linebreak
\arrow{l}{} P''_2 \setminus \cala_\calh$ as $l \notin \cala_\calh$, we observe that from $P'_2 \in
\reach(P_2)$ and $P_2 \in \mathrm{SBrSNNI}$ it follows that $P'_2 \setminus \cala_\calh \wbis{\rm b} P'_2 \,
/ \, \cala_\calh$, so that $P'_2 \setminus \cala_\calh \wbis{\rm b} P'_2 \, / \, \cala_\calh \wbis{\rm b}
P'_1 \setminus \cala_\calh$, i.e., $P'_2 \setminus \cala_\calh \wbis{\rm b} P'_1 \setminus \cala_\calh$.
Consequently, since $l \neq \tau$ there exist $\bar{P}'_1$ and $P''_1$ such that $P'_1 \setminus \cala_\calh
\warrow{\tau^{*}}{} \bar{P}'_1 \setminus \cala_\calh \linebreak \arrow{l}{} P''_1 \setminus \cala_\calh$
with $P'_2 \setminus \cala_\calh \wbis{\rm b} \bar{P}'_1 \setminus \cala_\calh$ and $P''_2 \setminus
\cala_\calh \wbis{\rm b} P''_1 \setminus \cala_\calh$. Therefore, $(((P'_2 \pco{L} Q) \, / \, L) \setminus
\cala_\calh, \bar{P}'_1 \setminus \cala_\calh) \in \calb$ -- because $\bar{P}'_1 \in \reach(P_1)$, $P'_2 \in
\reach(P_2)$, and $\bar{P}'_1 \setminus \cala_\calh \wbis{\rm b} P'_2 \, / \, \cala_\calh$ as $P_2 \in
\mathrm{SBrSNNI}$ -- and $(((P''_2 \pco{L} Q) \, / \, L) \setminus \cala_\calh, P''_1 \setminus \cala_\calh)
\in \calb$ -- because $P''_1 \in \reach(P_1)$, $P''_2 \in \reach(P_2)$, and $P''_1 \setminus \cala_\calh
\wbis{\rm b} P''_2 \, / \, \cala_\calh$ as $P_2 \in \mathrm{SBrSNNI}$.
 
\item If $((P'_2 \pco{L} Q) \, / \, L) \setminus \cala_\calh \arrow{\tau}{} ((P''_2 \pco{L} Q) \, / \, L)
\setminus \cala_\calh$ with $P'_2 \arrow{\tau}{} P''_2$ so that $P'_2 \setminus \cala_\calh \linebreak
\arrow{\tau}{} P''_2 \setminus \cala_\calh$ as $\tau \notin \cala_\calh$, we observe that from $P'_2 \in
\reach(P_2)$ and $P_2 \in \mathrm{SBrSNNI}$ it follows that $P'_2 \setminus \cala_\calh \wbis{\rm b} P'_2 \,
/ \, \cala_\calh$, so that $P'_2 \setminus \cala_\calh \wbis{\rm b} P'_2 \, / \, \cala_\calh \wbis{\rm b}
P'_1 \setminus \cala_\calh$, i.e., $P'_2 \setminus \cala_\calh \wbis{\rm b} P'_1 \setminus \cala_\calh$.
There are two subcases:

					\begin{itemize}

\item If $P''_2 \setminus \cala_\calh \wbis{\rm b} P'_1 \setminus \cala_\calh$ then $P'_1 \setminus
\cala_\calh$ is allowed to stay idle with $(((P''_2 \pco{L} Q) \, / \, L) \setminus \cala_\calh, \linebreak
P'_1 \setminus \cala_\calh) \in \calb$ because $P'_1 \in \reach(P_1)$, $P''_2 \in \reach(P_2)$, and $P'_1
\setminus \cala_\calh \wbis{\rm b} P''_2 \, / \, \cala_\calh$ \linebreak as $P_2 \in \mathrm{SBrSNNI}$.

\item If $P''_2 \setminus \cala_\calh \not\wbis{\rm b} P'_1 \setminus \cala_\calh$ then there exist
$\bar{P}'_1$ and $P''_1$ such that $P'_1 \setminus \cala_\calh \warrow{\tau^{*}}{} \bar{P}'_1 \setminus
\cala_\calh \linebreak \arrow{\tau}{} P''_1 \setminus \cala_\calh$ with $P'_2 \setminus \cala_\calh
\wbis{\rm b} \bar{P}'_1 \setminus \cala_\calh$ and $P''_2 \setminus \cala_\calh \wbis{\rm b} P''_1 \setminus
\cala_\calh$. Therefore, $(((P'_2 \pco{L} Q) \, / \, L) \setminus \cala_\calh, \bar{P}'_1 \setminus
\cala_\calh) \in \calb$ -- because $\bar{P}'_1 \in \reach(P_1)$, $P'_2 \in \reach(P_2)$, and $\bar{P}'_1
\setminus \cala_\calh \wbis{\rm b} P'_2 \, / \, \cala_\calh$ as $P_2 \in \mathrm{SBrSNNI}$ -- and $(((P''_2
\pco{L} Q) \, / \, L) \setminus \cala_\calh, P''_1 \setminus \cala_\calh) \in \calb$ \linebreak -- because
$P''_1 \in \reach(P_1)$, $P''_2 \in \reach(P_2)$, and $P''_1 \setminus \cala_\calh \wbis{\rm b} P''_2 \, /
\, \cala_\calh$ \linebreak as $P_2 \in \mathrm{SBrSNNI}$.

					\end{itemize}

\item If $((P'_2 \pco{L} Q) \, / \, L) \setminus \cala_\calh \arrow{\tau}{} ((P'_2 \pco{L} Q') \, / \, L)
\setminus \cala_\calh$ with $Q \arrow{\tau}{} Q'$, then trivially \linebreak $(((P'_2 \pco{L} Q') \, / \, L)
\setminus \cala_\calh, P'_1 \setminus \cala_\calh) \in \calb$.

\item If $((P'_2 \pco{L} Q) \, / \, L) \setminus \cala_\calh \arrow{\tau}{} ((P''_2 \pco{L} Q' \, / \, L)
\setminus \cala_\calh)$ with $P'_2 \arrow{h}{} P''_2$ -- so that $P'_2 \, / \, \cala_\calh \linebreak
\arrow{\tau}{} P''_2 \, / \, \cala_\calh$ as $h \in \cala_\calh$ -- and $Q \arrow{h}{} Q'$ for $h \in L$, we
observe that from $P'_2, P''_2 \in \reach(P_2)$ and $P_2 \in \mathrm{SBrSNNI}$ it follows that $P'_2
\setminus \cala_\calh \wbis{\rm b} P'_2 \, / \, \cala_\calh$ and $P''_2 \setminus \cala_\calh \wbis{\rm b}
P''_2 \, / \, \cala_\calh$, so that $P'_2 \setminus \cala_\calh \arrow{\tau}{} P''_2 \setminus \cala_\calh$
and $P'_2 \setminus \cala_\calh \wbis{\rm b} P'_2 \, / \, \cala_\calh \wbis{\rm b} P'_1 \setminus
\cala_\calh$, i.e., $P'_2 \setminus \cala_\calh \wbis{\rm b} P'_1 \setminus \cala_\calh$. There are two
subcases:

					\begin{itemize}

\item If $P''_2 \setminus \cala_\calh \wbis{\rm b} P'_1 \setminus \cala_\calh$ then $P'_1 \setminus
\cala_\calh$ is allowed to stay idle with $(((P''_2 \pco{L} Q') \, / \, L) \setminus \cala_\calh, \linebreak
P'_1 \setminus \cala_\calh) \in \calb$ because $P'_1 \in \reach(P_1)$, $P''_2 \in \reach(P_2)$, and $P'_1
\setminus \cala_\calh \wbis{\rm b} P''_2 \, / \, \cala_\calh$ \linebreak as $P_2 \in \mathrm{SBrSNNI}$.

\item If $P''_2 \setminus \cala_\calh \not\wbis{\rm b} P'_1 \setminus \cala_\calh$ then there exist
$\bar{P}'_1$ and $P''_1$ such that $P'_1 \setminus \cala_\calh \warrow{\tau^{*}}{} \bar{P}'_1 \setminus
\cala_\calh \linebreak \arrow{\tau}{} P''_1 \setminus \cala_\calh$ with $P'_2 \setminus \cala_\calh
\wbis{\rm b} \bar{P}'_1 \setminus \cala_\calh$ and $P''_2 \setminus \cala_\calh \wbis{\rm b} P''_1 \setminus
\cala_\calh$. Therefore, $(((P'_2 \pco{L} Q) \, / \, L) \setminus \cala_\calh, \bar{P}'_1 \setminus
\cala_\calh) \in \calb$ -- because $\bar{P}'_1 \in \reach(P_1)$, $P'_2 \in \reach(P_2)$, and $\bar{P}'_1
\setminus \cala_\calh \wbis{\rm b} P'_2 \, / \, \cala_\calh$ as $P_2 \in \mathrm{SBrSNNI}$ -- and $(((P''_2
\pco{L} Q') \, / \, L) \setminus \cala_\calh, P''_1 \setminus \cala_\calh) \in \calb$ -- because $P''_1 \in
\reach(P_1)$, $P''_2 \in \reach(P_2)$, and $P''_1 \setminus \cala_\calh \wbis{\rm b} P''_2 \, / \,
\cala_\calh$ as $P_2 \in \mathrm{SBrSNNI}$.
\qedhere

					\end{itemize}

				\end{itemize}

			\end{enumerate}

		\end{proof}

	\end{lem}

	\begin{thm}\label{thm:branching_taxonomy_1}

$\mathrm{SBrNDC} \subset \mathrm{SBrSNNI} = \mathrm{P\_BrNDC} \subset \mathrm{BrNDC} \subset
\mathrm{BrSNNI}$.

		\begin{proof}
Let us examine each relationship separately:

			\begin{itemize}

\item SBrNDC $\subset$ SBrSNNI.
Given $P \in \mathrm{SBrNDC}$, the result follows by proving that the symmetric relation $\calb = \{ (P'
\setminus \cala_\calh, P' \, / \, \cala_\calh), (P' \, / \, \cala_\calh, P' \setminus \cala_\calh) \mid P'
\in \reach(P) \}$ is a branching bisimulation up to $\wbis{\rm b}$. Starting from $P' \setminus \cala_\calh$
and $P' \, / \, \cala_\calh$ related by $\calb$, \linebreak in the up-to branching bisimulation game there
are three cases based on the operational semantic rules in Table~\ref{tab:op_sem}. In the first case, it is
$P' \setminus \cala_\calh$ to move first:

				\begin{itemize}

\item If $P' \setminus \cala_\calh \warrow{\tau^{*}}{} \bar{P'} \setminus \cala_\calh \arrow{a}{} P''
\setminus \cala_\calh$ with $\bar{P'} \arrow{a}{} P''$ and $a \in \cala_\call \cup \{ \tau \}$, then $P' /
\cala_\calh \warrow{\tau^{*}}{}$ \linebreak $\bar{P'} / \cala_\calh \arrow{a}{} P'' / \cala_\calh$ as $\tau,
a \notin \cala_\calh$, with $(\bar{P'} \setminus \cala_\calh, \bar{P'} \, / \, \cala_\calh) \in \calb$ and
$(P'' \setminus \cala_\calh, P'' \, / \, \cala_\calh) \in \calb$ so that $\bar{P'} \setminus \cala_\calh
\wbis{\rm b} \bar{P'} \setminus \cala_\calh \; \calb \; \bar{P'} \, / \, \cala_\calh \wbis{\rm b} \bar{P'}
\, / \, \cala_\calh$ and $P'' \setminus \cala_\calh \wbis{\rm b} P'' \setminus \cala_\calh \linebreak \calb
\; P'' \, / \, \cala_\calh \wbis{\rm b} P'' \, / \, \cala_\calh$.

				\end{itemize}

\noindent
In the other two cases, instead, it is $P' \, / \, \cala_\calh$ to move first (note that possible
$\tau$-transitions arising from high actions in $P'$ can no longer be executed when starting from $P'
\setminus \cala_\calh$):

				\begin{itemize}

\item If $P' \, / \, \cala_\calh \warrow{\tau^{*}}{} \bar{P'} \, / \, \cala_\calh \arrow{a}{} P'' \, / \,
\cala_\calh$ with $\bar{P'} \arrow{a}{} P''$ and $a \in \cala_\call \cup \{ \tau \}$, then $\bar{P'}
\setminus \cala_\calh \linebreak \arrow{a}{} P'' \setminus \cala_\calh$ as $a \notin \cala_\calh$. Since $P'
\, / \, \cala_\calh \warrow{\tau^{*}}{} \bar{P'} \, / \, \cala_\calh$ implies $P' \setminus \cala_\calh
\warrow{\tau^{*}}{} \hat{P'} \setminus \cala_\calh$ with $\bar{P'} \setminus \cala_\calh \wbis{\rm b}
\hat{P'} \setminus \cala_\calh$ by virtue of Lemma~\ref{lem:branching_taxonomy}(1), from $\bar{P'} \setminus
\cala_\calh \arrow{a}{} P'' \setminus \cala_\calh$ it follows that:

					\begin{itemize}

\item Either $a = \tau$ and $P'' \setminus \cala_\calh \wbis{\rm b} \hat{P'} \setminus \cala_\calh$, hence
$\bar{P'} \setminus \cala_\calh \wbis{\rm b} P'' \setminus \cala_\calh$ as $\wbis{\rm b}$ is symmetric and
transitive. From Lemma~\ref{lem:branching_taxonomy}(2) it then follows that $\bar{P'} \, / \, \cala_\calh
\wbis{\rm b}{} P'' \, / \, \cala_\calh$ because $\bar{P'}, P'' \in \mathrm{SBrNDC}$ as $\bar{P'}, P'' \in
\reach(P)$ and $P \in \mathrm{SBrNDC}$. Therefore $P' \setminus \cala_\calh$ can stay idle in the up-to
branching bisimulation game.

\item Or $\hat{P'} \setminus \cala_\calh \warrow{\tau^{*}}{} \hat{P}'' \setminus \cala_\calh \arrow{a}{}
\hat{P}''' \setminus \cala_\calh$ with $\bar{P'} \setminus \cala_\calh \wbis{\rm b} \hat{P}'' \setminus
\cala_\calh$ and $P'' \setminus \cala_\calh \wbis{\rm b} \hat{P}''' \setminus \cala_\calh$. From $P'
\setminus \cala_\calh \warrow{\tau^{*}}{} \hat{P'} \setminus \cala_\calh$ it follows that $P' \setminus
\cala_\calh \warrow{\tau ^{*}}{} \hat{P}'' \setminus \cala_\calh \arrow{a}{} \hat{P}''' \setminus
\cala_\calh$ with $\bar{P'} \, / \, \cala_\calh \wbis{\rm b} \bar{P'} \, / \, \cala_\calh \; \calb \;
\bar{P'} \setminus \cala_\calh \wbis{\rm b} \hat{P}'' \setminus \cala_\calh$ and $P'' \, / \, \cala_\calh
\wbis{\rm b} P'' \, / \, \cala_\calh \; \calb \; P'' \setminus \cala_\calh \wbis{\rm b} \hat{P}''' \setminus
\cala_\calh$.

					\end{itemize}

\item If $P' \, / \, \cala_\calh \warrow{\tau^{*}}{} \bar{P'} \, / \, \cala_\calh \arrow{\tau}{} P'' \, / \,
\cala_\calh$ with $\bar{P'} \arrow{h}{} P''$, we observe that $\bar{P'} \setminus \cala_\calh$ cannot
perform any $h$-action as $h \in \cala_{\calh}$, nor we know whether it can perform a $\tau$-action --
moreover $(P'' \, / \, \cala_\calh, P' \backslash \cala_\calh) \notin \calb$ when $P''$ is different from
$P'$, hence the need of resorting to the up-to technique. However, from $\bar{P'} \in \reach(P)$ and $P \in
\mathrm{SBrNDC}$ \linebreak it follows that $\bar{P'} \setminus \cala_\calh \wbis{\rm b} P'' \setminus
\cala_\calh$, hence $\bar{P'} \, / \, \cala_\calh \wbis{\rm b} P'' \, / \, \cala_\calh$ by virtue of
Lemma~\ref{lem:branching_taxonomy}(2) because $\bar{P'}, P'' \in \mathrm{SBrNDC}$. Therefore $P' \setminus
\cala_\calh$ can stay idle in the up-to branching bisimulation game.

				\end{itemize}

\item SBrSNNI = P\_BrNDC.
We first prove that $\mathrm{P\_BrNDC} \subseteq \mathrm{SBrSNNI}$. If $P \in \mathrm{P\_BrNDC}$ then $P'
\in \mathrm{BrNDC}$ for every $P' \in \reach(P)$. Since $\mathrm{BrNDC} \subset \mathrm{BrSNNI}$ as we will
show in the last part of the proof of this theorem, $P' \in \mathrm{BrSNNI}$ for every $P' \in \reach(P)$,
i.e., $P \in \mathrm{SBrSNNI}$. \\
The fact that $\mathrm{SBrSNNI} \subseteq \mathrm{P\_BrNDC}$ follows from
Lemma~\ref{lem:branching_taxonomy}(3) by taking $P'_1$ identical to~$P'_2$ and both reachable from $P \in
\mathrm{SBrSNNI}$.

\item SBrSNNI $\subset$ BrNDC.
If $P \in \mathrm{SBrSNNI} = \mathrm{P\_BrNDC}$ then it immediately follows that $P \in \mathrm{BrNDC}$.

\item BrNDC $\subset$ BrSNNI.
If $P \in \mathrm{BrNDC}$, i.e., $P \setminus \cala_\calh \wbis{\rm b} (P \pco{L} Q) \, / \, L) \setminus
\cala_\calh$ for all $Q \in \procs$ such that every $Q' \in \reach(Q)$ has only actions in $\cala_\calh$ and
for all $L \subseteq \cala_\calh$, then we can consider in particular $\hat{Q}$ capable of stepwise
mimicking the high-level behavior of $P$, in the sense that $\hat{Q}$ is able to synchronize with all the
high-level actions executed by $P$ and its reachable processes, along with $\hat{L} = \cala_\calh$.  As a
consequence $(P \pco{\hat{L}} \hat{Q}) \, / \, \hat{L}) \setminus \cala_\calh$ \linebreak is isomorphic to
$P \, / \, \cala_\calh$, hence $P \setminus \cala_\calh \wbis{\rm b} P \, / \, \cala_\calh$, i.e., $P \in
\mathrm{BrSNNI}$.
\qedhere

			\end{itemize}

		\end{proof}

	\end{thm}

\noindent
All the inclusions above are strict as we now show:

	\begin{itemize}

\item The process $\tau \, . \, l \, . \, \nil + l \, . \, l \, . \, \nil + h \, . \, l \, . \, \nil$ is
SBrSNNI (resp.\ P\_BrNDC) because $(\tau \, . \, l \, . \, \nil + l \, . \, l \, . \, \nil + h \, . \, l \,
. \, \nil) \setminus \{ h \} \wbis{\rm b} (\tau \, . \, l \, . \, \nil + l \, . \, l \, . \, \nil + h \, .
\, l \, . \, \nil) \, / \, \{ h \}$ and action $h$ is enabled only by the initial process so every reachable
process is BrSNNI (resp.\ BrNDC). It is not SBrNDC because the low-level view of the process reached after
action $h$, i.e., $(l \, . \, \nil) \setminus \{ h \}$, is not branching bisimilar to $(\tau \, . \, l \, .
\, \nil + l \, . \, l \, . \, \nil + h \, . \, l \, . \, \nil) \setminus \{ h \}$.

\item The process $l \, . \, \nil + l \, . \, l \, . \, \nil + l \, . \, h \, . \, l \, . \, \nil$ is BrNDC
because, whether there are synchronizations with high-level actions or not, the overall process can always
perform either an $l$-action or a sequence of two $l$-actions without incurring any problematic branching.
The process is not SBrSNNI (resp.\ P\_BrNDC) because the reachable process $h \, . \, l \, . \, \nil$ is not
BrSNNI \linebreak (resp.\ BrNDC).

\item The process $l \, . \, \nil + h \, . \, h \, . \, l \, . \, \nil$ is BrSNNI as $(l\, . \, \nil + h \,
. \, h \, . \, l \, . \, \nil) \setminus \{ h \} \wbis{\rm b} (l \, . \, \nil + h \, . \, h \, .\, l \, .
\, \nil) \, / \, \{ h \} $. It is not BrNDC due to $(((l \, . \, \nil + h \, . \, h \, . \, l \, . \, \nil)
\pco{\{ h \}} (h \, . \, \nil)) \, / \, \{ h \}) \setminus \{ h \} \not\wbis{\rm b} (l \, . \, \nil + h \, .
\, h \, . \, l \, . \, \nil) \setminus \{ h \}$ because the former behaves as $l \, . \, \nil + \tau \, . \,
\nil$ while the latter behaves as $l \, . \, \nil$.

	\end{itemize}
\noindent 
We further observe that each of the $\wbis{\rm b}$-based noninterference properties listed in
Definition~\ref{def:branching_bisim_properties} implies the corresponding $\wbis{}$-based noninterference
property listed in Definition~\ref{def:weak_bisim_properties}. This is simply due to the fact that
$\wbis{\rm b}$ is finer than $\wbis{}$~\cite{GW96}.

	\begin{thm}\label{thm:branching_taxonomy_2}

The following inclusions hold:

		\begin{enumerate}

\item $\mathrm{BrSNNI} \subset \mathrm{BSNNI}$.

\item $\mathrm{BrNDC} \subset \mathrm{BNDC}$.

\item $\mathrm{SBrSNNI} \subset \mathrm{SBSNNI}$.

\item $\mathrm{P\_BrNDC} \subset \mathrm{P\_BNDC}$.

\item $\mathrm{SBrNDC} \subset \mathrm{SBNDC}$.	
\fullbox

		\end{enumerate}

	\end{thm}
\noindent 
All the inclusions above are strict by virtue of the following result; for an example of $P_1$ and $P_2$
below, see Figure~\ref{fig:wb_brb_cex}.

	 \begin{thm}\label{thm:branching_taxonomy_3}

Let $P_1, P_2 \in \procs$ be such that $P_1 \wbis{} P_2$ but $P_1 \not\wbis{\rm b} P_2$. If no high-level
actions occur in $P_1$ and $P_2$, then $Q \in \{ P_1 + h \, . \, P_2, P_2 + h \, . \, P_1 \}$ is such that:

		\begin{enumerate}

\item $Q \in \mathrm{BSNNI}$ but $Q \notin \mathrm{BrSNNI}$.

\item $Q \in \mathrm{BNDC}$ but $Q \notin \mathrm{BrNDC}$.

\item $Q \in \mathrm{SBSNNI}$ but $Q \notin \mathrm{SBrSNNI}$.

\item $Q \in \mathrm{P\_BNDC}$ but $Q \notin \mathrm{P\_BrNDC}$.

\item $Q \in \mathrm{SBNDC}$ but $Q \notin \mathrm{SBrNDC}$.

		\end{enumerate}

		\begin{proof}
Let $Q$ be $P_1 + h \, . \, P_2$ (the proof is similar for $Q$ equal to $P_2 + h \, . \, P_1$) and observe
that no high-level actions occur in every process reachable from $Q$ except $Q$ itself:

			\begin{enumerate}

\item Let $\calb$ be a weak bisimulation witnessing $P_1 \wbis{} P_2$. Then $Q \in \mathrm{BSNNI}$ because
the symmetric relation $\calb' = \calb \cup \{ (Q \setminus \cala_\calh, Q \, / \, \cala_\calh), (Q \, / \,
\cala_\calh, Q \setminus \cala_\calh) \}$ turns out to be a weak bisimulation too. The only interesting case
is the one where $Q \, / \, \cala_\calh$, which is isomorphic to $P_1 + \tau \, . \, P_2$, performs a
$\tau$-action toward $P_2 \, / \, \cala_\calh$, which is isomorphic to $P_2$. \linebreak In that case $Q
\setminus \cala_\calh$, which is isomorphic to $P_1$, can respond by staying idle, because $(P_2, P_1) \in
\calb$ and hence $(P_2, P_1) \in \calb'$. \\
On the other hand, $Q \notin \mathrm{BrSNNI}$ because $P_2 \not\wbis{\rm b} P_1$ in the same situation as
before.

\item Since $Q \in \mathrm{BSNNI}$ and no high-level actions occur in every process reachable from $Q$ other
than $Q$, it holds that $Q \in \mathrm{SBSNNI}$ and hence $Q \in \mathrm{BNDC}$ by virtue of
Theorem~\ref{thm:weak_bisim_taxonomy}. \\
On the other hand, from $Q \notin \mathrm{BrSNNI}$ it follows that $Q \notin \mathrm{BrNDC}$ by virtue of
\linebreak Theorem~\ref{thm:branching_taxonomy_1}.

\item We already know from the previous case that $Q \in \mathrm{SBSNNI}$. \\ On the other hand, from $Q
\notin \mathrm{BrSNNI}$ it follows that $Q \notin \mathrm{SBrSNNI}$ by virtue of
Theorem~\ref{thm:branching_taxonomy_1}.

\item A straightforward consequence of P\_BNDC = SBSNNI (Theorem~\ref{thm:weak_bisim_taxonomy}) and P\_BrNDC
= SBrSNNI (Theorem~\ref{thm:branching_taxonomy_1}).

\item Since the only high-level action occurring in $Q$ is $h$, in the proof of $Q \in \mathrm{SBNDC}$ the
only interesting case is the transition $Q \arrow{h}{} P_2$, for which it holds that $Q \setminus
\cala_\calh \wbis{} P_2 \setminus \cala_\calh$ because the former is isomorphic to $P_1$, the latter is
isomorphic to $P_2$, and $P_1 \wbis{} P_2$. \\
On the other hand, $Q \notin \mathrm{SBrNDC}$ because $P_1 \not\wbis{\rm b} P_2$ in the same situation as
before.
\qedhere

			\end{enumerate}

		\end{proof}

	\end{thm}
\noindent 
An alternative strategy to explore the differences between $\wbis{}$ and $\wbis{\rm b}$ with respect to
B/BrSNNI and SB/BrSNNI consists of considering the two $\tau$-axioms $\tau \, . \, x + x = \tau \, . \, x$
and $a \, . \, (\tau \, . \, x + y) + a \, . \, x = a \, . \, (\tau \, . \, x + y)$ for
$\wbis{}$~\cite{Mil89a}. The strategy is inspired by the initial remarks in~\cite{GW96}, where it is noted
that the two aforementioned axioms are not valid for $\wbis{\rm b}$ and are responsible for the lack of
distinguishing power of $\wbis{}$ over $\tau$-branching processes. For each axiom, the strategy is based on
constructing a pair of new processes from the ones equated in the axiom, such that they are weakly bisimilar
by construction but not branching bisimilar. Then from this pair of processes we define a new process $P$
such that $P \setminus \cala_\calh$ and $P \, / \, \cala_\calh$ are isomorphic to the constructed processes.
 
	\begin{fact}\label{thm:cex_tau_axiom_2}	

From $\tau \, . \, x + x = \tau \, . \, x$ it is possible to construct $P \in \procs$ such that $P \in
\mathrm{BSNNI}$ but $P \notin \mathrm{BrSNNI}$ and $P \in \mathrm{SBSNNI}$ but $P \notin \mathrm{SBrSNNI}$.

		\begin{proof}
In $\tau \, . \, x + x  = \tau \, . \, x$ let us instantiate $x$ as $\tau \, . \, l_1 \, . \, \nil + \tau \,
. \, l_2 \, . \, \nil$ and then add $+ \, l_3 \, . \, \nil$ to both sides of the equation thus obtaining
$\tau \, . \, (\tau \, . \, l_1 \, . \, \nil + \tau \, . \, l_2 \, . \, \nil) + (\tau \, . \, l_1 \, . \,
\nil + \tau \, . \, l_2 \, . \, \nil) + l_3 \, . \, \nil = \tau \, . \, (\tau \, . \, l_1 \, . \, \nil +
\tau \, . \, l_2 \, . \, \nil) + l_3 \, . \, \nil$, which are related by weak bisimilarity but not by
branching bisimilarity. Now let us define process $P$ as $\tau \, . \, (\tau \, . \, l_1 \, . \, \nil + \tau
\, . \, l_2 \, . \, \nil) + (h \, . \, l_1 \, . \, \nil + h \, . \, l_2 \, . \, \nil) + l_3 \, . \, \nil$,
for which it holds that $P / \cala_\calh$ and $P \setminus \cala_\calh$ are isomorphic to the two sides of
the equation, respectively. By construction, it immediately follows that $P$ is BSNNI but not BrSNNI. \\
Since the only high-level action is performed by $P$ itself, which is BSNNI, for every other process $P'$
reachable from $P$ it holds that $P' \setminus \cala_\calh$ is isomorphic to $P' \, / \, \cala_\calh$, hence
\linebreak $P \in \mathrm{SBSNNI}$ but $P \notin \mathrm{SBrSNNI}$.
\qedhere

		\end{proof}

	\end{fact}

	\begin{fact}\label{thm:cex_tau_axiom_3}	

From $a \, . \, (\tau \, . \, x + y) + a \, . \, x = a \, . \, (\tau \, . \, x + y) $ it is possible to
construct $P \in \procs$ such that $P \in \mathrm{BSNNI}$ but $P \notin \mathrm{BrSNNI}$ and $P \in
\mathrm{SBSNNI}$ but $P \notin \mathrm{SBrSNNI}$.

		\begin{proof}
In $a \, . \, (\tau \, . \, x + y) + a \, . \, x = a \, . \, (\tau \, . \, x + y) $ let us instantiate $a$
as $\tau$, $x$ as $l_1 \, . \, \nil$, and $y$ as $l_2 \, . \, \nil$ and then add $+ \, l_3 \, . \, \nil$ to
both sides of the equation thus obtaining $\tau \, . \, (\tau \, . \, l_1 \, . \, \nil + l_2 \, . \, \nil) +
\tau \, . \, l_1 \, . \, \nil + l_3 \, . \, \nil = \tau \, . \, (\tau \, . \, l_1 \, . \, \nil + l_2 \, . \,
\nil) + l_3 \, . \, \nil$, which are related by weak bisimilarity but not by branching bisimilarity. Now let
us define process $P$ as $\tau \, . \, (\tau \, . \, l_1 \, . \, \nil + l_2 \, . \, \nil) + h \, . \, l_1 \,
. \, \nil + l_3 \, . \, \nil$, for which it holds that $P \, / \, \cala_\calh$ and $P \setminus \cala_\calh$
are isomorphic to the two sides of the equation, respectively. By construction, it immediately follows that
$P$ is BSNNI but not BrSNNI. \\
Since the only high-level action is performed by $P$ itself, which is BSNNI, for every other process $P'$
reachable from $P$ it holds that $P' \setminus \cala_\calh$ is isomorphic to $P' \, / \, \cala_\calh$, hence
\linebreak $P \in \mathrm{SBSNNI}$ but $P \notin \mathrm{SBrSNNI}$.
\qedhere

		\end{proof}

	\end{fact}

	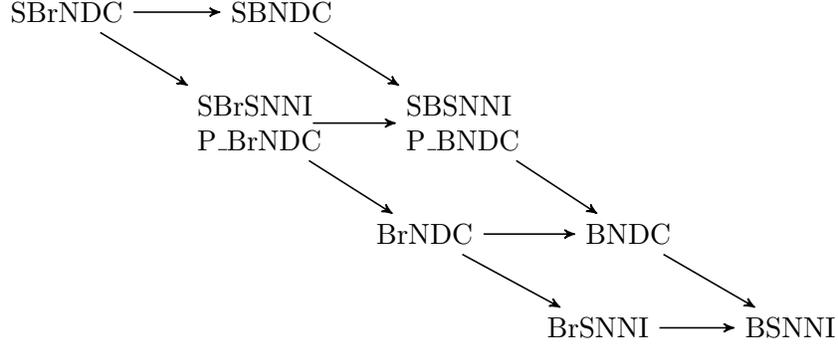
\begin{figure}[t]

\begin{center}
\begin{tikzpicture}[modal]

\node (bsnni)                                                         {BSNNI};
\node (bndc)    [above left = of bsnni]                               {BNDC};
\node (pbsnni)  [above left = of bndc, text width=1.4cm, align=left]  {SBSNNI \\ P\_BNDC};
\node (sbndc)   [above left = of pbsnni]                              {SBNDC};
\node (brsnni)  [left = of bsnni]                                     {BrSNNI};
\node (brndc)   [above left = of brsnni]                              {BrNDC};
\node (pbrsnni) [above left = of brndc, text width=1.4cm, align=left] {SBrSNNI \\ P\_BrNDC};
\node (sbrndc)  [above left = of pbrsnni]                             {SBrNDC};

\path[<-] (bndc)    edge (pbsnni);
\path[<-] (bsnni)   edge (bndc);
\path[<-] (bsnni)   edge (brsnni);
\path[<-] (pbsnni)  edge (sbndc);
\path[<-] (pbsnni)  edge (pbrsnni);
\path[<-] (brndc)   edge (pbrsnni);
\path[<-] (brsnni)  edge (brndc);
\path[<-] (bndc)    edge (brndc);
\path[<-] (sbndc)   edge (sbrndc);
\path[<-] (pbrsnni) edge (sbrndc);

\end{tikzpicture}
\end{center}

\caption{Taxonomy of security properties based on weak and branching bisimilarities}
\label{fig:taxon_weak_branching}

	\end{figure}
\noindent 
Based on the results in Theorems~\ref{thm:weak_bisim_taxonomy}, \ref{thm:branching_taxonomy_1},
and~\ref{thm:branching_taxonomy_2}, the diagram in Figure~\ref{fig:taxon_weak_branching} summarizes the
inclusions among the various noninterference properties, where $\calp \rightarrow \calq$ means that $\calp$
is strictly included in $\calq$. The missing arrows in the diagram, witnessing incomparability, are
justified by the following counterexamples:

	\begin{itemize}

\item SBNDC vs.\ SBrSNNI. The process $\tau \, . \, l \, . \, \nil + l \, . \, l \, . \, \nil + h \, . \, l
\, . \, \nil$ is BrSNNI as $\tau \, . \, l \, . \, \nil + l \, . \, l \, . \, \nil \wbis{\rm b} \tau \, . \,
l \, . \, \nil + l \, . \, l \, . \, \nil + \tau \, . \, l \, . \, \nil$. It is also SBrSNNI because every
reachable process does not enable any more high-level actions. However, it is not SBNDC, because after
executing the high-level action $h$ it can perform a single $l$-action, while the original process with the
restriction on high-level actions can go along a path where it performs two $l$-actions. On the other hand,
the process $Q$ mentioned in Theorem~\ref{thm:branching_taxonomy_3} is SBNDC but neither BrSNNI nor SBrSNNI.

\item SBSNNI vs.\ BrNDC. The process $l \, . \, h \, . \, l \, . \, \nil + l \, . \, \nil + l \, . \, l \, .
\, \nil$ is BrSNNI as $l \, . \, \nil + l \, . \, \nil + l \, . \, l \, . \, \nil \wbis{\rm b} l \, . \,
\tau \, . \, l \, . \, \nil + l \, . \, \nil + l \, . \, l \, . \, \nil$. In particular, the subprocesses $l
\, . \, \tau \, . \, l \, . \, \nil$ and $l \, . \, l \, . \, \nil$ are equated by virtue of the other axiom
of weak bisimilarity, $a \, . \, \tau \, . \, x = a \, . \, x$, which holds also for branching bisimilarity.
The same process is BrNDC too as it includes only one high-level action, hence the only possible high-level
strategy coincides with the check conducted by BrSNNI. However, the process is not SBSNNI because of the
reachable process $h \, . \, l \, . \, \nil$, which is not BSNNI. On the other hand, the process $Q$
mentioned in Theorem~\ref{thm:branching_taxonomy_3} \linebreak is SBSNNI but not BrSNNI and, therefore,
cannot be BrNDC.

\item BNDC vs.\ BrSNNI. The process $l \, . \, \nil + h_1 \, . \, h_2 \, . \, l \, . \, \nil$ is not BNDC
(see Section~\ref{sec:weak_bisim_properties}), but it is BrSNNI as $l \, . \, \nil \wbis{\rm b} l \, . \,
\nil + \tau \, . \, \tau \, . \, l \, . \, \nil$. In contrast, the process $Q$ mentioned in
Theorem~\ref{thm:branching_taxonomy_3} is both BSNNI and BNDC, but not BrSNNI.

	\end{itemize}
\noindent 
It is worth noting that the strongest property based on weak bisimilarity (SBNDC) and the weakest property
based on branching bisimilarity (BrSNNI) are incomparable too. The former is a very restrictive property
because it requires a local check every time a high-level action is performed, while the latter requires a
check only on the initial state. On the other hand, as shown in Theorem~\ref{thm:branching_taxonomy_3} it is
very easy to construct processes that are secure under properties based on $\wbis{}$ but not on $\wbis{\rm
b}$ due to the minimal number of high-level actions in $Q$.

%
%
\section{Noninterference in Reversible Processes}
\label{sec:bf}
%
%

As anticipated, we use reversible computing to motivate the study of branching-bisimilarity-based
noninterference properties. To this aim, we now recall from~\cite{DMV90} back-and-forth bisimilarity and its
relationship with standard forward-only bisimilarity.

An LTS represents a reversible process if each of its transitions is seen as bidirectional. This means that
the action labeling every transition can be undone and then redone. When going backward, it is of paramount
importance to respect causality. While this is straightforward for sequential processes, it is not obvious
for concurrent ones, because the last performed action is the first one to be undone but this action may not
necessarily be identifiable uniquely in the presence of concurrency.

Consider for example a process that can perform action $a$ in parallel with action $b$. This process can be
represented as a diamond-like LTS where from the initial state an $a$-transition and a $b$-transition
depart, which are respectively followed by a $b$-transition and an $a$-transition both of which reach the
final state. Suppose that action $a$ completes before action $b$, so that the $a$-transition is executed
before the $b$-transition. Once in the final state, either the $b$-transition is undone before the
$a$-transition, or the $a$-transition is undone before the $b$-transition. Both options are causally
consistent, as $a$ and $b$ are independent of each other, but only the former is history preserving too.

The history-preserving option is the one that was addressed in~\cite{DMV90} in order to study reversible
processes in an interleaving setting. To accomplish this, strong and weak bisimulations were redefined as
binary relations between histories, formalized below as runs, instead of states. The resulting behavioral
equivalences are respectively called strong and weak back-and-forth bisimilarities in~\cite{DMV90}.

	\begin{defi}\label{def:path_run_comp_trans}

Let $(\cals, \cala_{\tau}, \! \arrow{}{} \!)$ be an LTS:

		\begin{itemize}

\item A sequence $\xi = s \arrow{a_{1}}{} s_{1} \arrow{a_{2}}{} s_{2} \dots s_{n - 1} \arrow{a_{n}}{} s_{n}$
is called a \emph{path} from state $s \in \cals$ of length $n \in \natns$, where we let $\first(\xi) = s$
and $\last(\xi) = s_{n}$. We denote by $\Path(s)$ the set of paths from state $s$, including the empty path
indicated with $\varepsilon$.

\item A pair $\rho = (s, \xi)$ is called a \emph{run} from state $s \in \cals$ iff $\xi \in \Path(s)$, in
which case we let $\pt(\rho) = \xi$, $\first(\rho) = \first(\xi)$, and $\last(\rho) = \last(\xi)$, with
$\first(\rho) = \last(\rho) = s$ when $\xi = \varepsilon$. We denote by $\Run(s)$ the set of runs from state
$s$.

\item Let $\rho = (s, \xi) \in \Run(s)$ and $\rho' = (s', \xi') \in \Run(s')$ for $s, s' \in \cals$:

			\begin{itemize}

\item Their composition $\rho \rho' = (s, \xi \xi') \in \Run(s)$ is defined iff $\last(\rho) =
\first(\rho')$.

\item We write $\rho \arrow{a}{} \rho'$ iff there exists $\bar{\rho} = (\bar{s}, \bar{s} \arrow{a}{} s')$
with $\bar{s} = \last(\rho)$ such that $\rho' = \rho \bar{\rho}$. 
\fullbox

			\end{itemize}

		\end{itemize}

	\end{defi}
\noindent 
In the behavioral equivalences of~\cite{DMV90}, for any given LTS the set $\calr$ of its runs is considered
in lieu of the set $\cals$ of its states. Using runs instead of just paths is convenient in the case of an
empty path so as to know the state under consideration. Given a pair of runs $(\rho_{1}, \rho_{2})$, in the
two definitions below the forward clauses consider outgoing transitions whereas the backward clauses
consider incoming transitions.

	\begin{defi}\label{def:bf_bisim}

Let $(\cals, \cala_{\tau}, \! \arrow{}{} \!)$ be an LTS and $s_{1}, s_{2} \in \cals$. We say that $s_{1}$
and~$s_{2}$ are \emph{strongly back-and-forth bisimilar}, written $s_{1} \sbis{\rm bf} s_{2}$, iff $((s_{1},
\varepsilon), (s_{2}, \varepsilon)) \in \calb$ for some strong back-and-forth bisimulation $\calb$. A
symmetric binary relation $\calb$ over $\calr$ is a \emph{strong back-and-forth bisimulation} iff, whenever
$(\rho_{1}, \rho_{2}) \in \calb$, then:

		\begin{itemize}

\item for each $\rho_{1} \arrow{a}{} \rho'_{1}$ there exists $\rho_{2} \arrow{a}{} \rho'_{2}$ such that
$(\rho'_{1}, \rho'_{2}) \in \calb$;

\item for each $\rho'_{1} \arrow{a}{} \rho_{1}$ there exists $\rho'_{2} \arrow{a}{} \rho_{2}$ such that
$(\rho'_{1}, \rho'_{2}) \in \calb$.
\fullbox

		\end{itemize}

	\end{defi}

	\begin{defi}\label{def:weak_bf_bisim}

Let $(\cals, \cala_{\tau}, \! \arrow{}{} \!)$ be an LTS and $s_{1}, s_{2} \in \cals$. We say that $s_{1}$
and~$s_{2}$ are \emph{weakly back-and-forth bisimilar}, written $s_{1} \wbis{\rm bf} s_{2}$, iff $((s_{1},
\varepsilon), (s_{2}, \varepsilon)) \in \calb$ for some weak back-and-forth bisimulation $\calb$. A
symmetric binary relation $\calb$ over $\calr$ is a \emph{weak back-and-forth bisimulation} iff, whenever
$(\rho_{1}, \rho_{2}) \in \calb$, then:

		\begin{itemize}

\item for each $\rho_{1} \arrow{\tau}{} \rho'_{1}$ there exists $\rho_{2} \warrow{\tau^{*}}{} \rho'_{2}$
such that $(\rho'_{1}, \rho'_{2}) \in \calb$;

\item for each $\rho'_{1} \arrow{\tau}{} \rho_{1}$ there exists $\rho'_{2} \warrow{\tau^{*}}{} \rho_{2}$
such that $(\rho'_{1}, \rho'_{2}) \in \calb$;

\item for each $\rho_{1} \arrow{a}{} \rho'_{1}$ with $a \in \cala$ there exists $\rho_{2}
\warrow{\tau^{*}}{} \! \arrow{a}{} \! \warrow{\tau^{*}}{} \rho'_{2}$ such that $(\rho'_{1}, \rho'_{2}) \in
\calb$;

\item for each $\rho'_{1} \arrow{a}{} \rho_{1}$ with $a \in \cala$ there exists $\rho'_{2}
\warrow{\tau^{*}}{} \! \arrow{a}{} \! \warrow{\tau^{*}}{} \rho_{2}$ such that $(\rho'_{1}, \rho'_{2}) \in
\calb$.
\fullbox

		\end{itemize}

	\end{defi}
\noindent 
In~\cite{DMV90} it was shown that strong back-and-forth bisimilarity coincides with strong bisimilarity.
Surprisingly, weak back-and-forth bisimilarity does not coincide with weak bisimilarity. Instead, it
coincides with branching bisimilarity. For example, in Figure~\ref{fig:wb_brb_cex} \linebreak it holds that
$s_{1} \not\wbis{\rm bf} s_{2}$ because in the forward direction $(s_{1}, \varepsilon) \arrow{a}{} (s_{1},
s_{1} \arrow{a}{} s'_{1})$ is matched by $(s_{2}, \varepsilon) \arrow{\tau}{} (s_{2}, s_{2} \arrow{\tau}{}
s'_{2}) \arrow{a}{} (s_{2}, s_{2} \arrow{\tau}{} s'_{2} \arrow{a}{} s''_{2})$, but then in the backward
direction $(s_{2}, s_{2} \arrow{\tau}{} s'_{2}) \arrow{a}{} (s_{2}, s_{2} \arrow{\tau}{} s'_{2} \arrow{a}{}
s''_{2})$ is not matched by $(s_{1}, \varepsilon) \arrow{a}{} (s_{1}, s_{1} \arrow{a}{} s'_{1})$ because
$(s_{1}, \varepsilon)$ has an outgoing $b$-transition whilst $(s_{2}, s_{2} \arrow{\tau}{} s'_{2})$ has not.

	\begin{thm}\label{thm:bf_bisim_eq_bisim}

Let $(\cals, \cala_{\tau}, \! \arrow{}{} \!)$ be an LTS and $s_{1}, s_{2} \in \cals$. Then: 

		\begin{itemize}

\item $s_{1} \sbis{\rm bf} s_{2}$ iff $s_{1} \sbis{} s_{2}$.

\item $s_{1} \wbis{\rm bf} s_{2}$ iff $s_{1} \wbis{\rm b} s_{2}$.
\fullbox

		\end{itemize}

	\end{thm}
\noindent 
As a consequence, the properties BrSNNI, BrNDC, SBrSNNI, P\_BrDNC, and SBrNDC do not change if $\wbis{\rm
b}$ is replaced by $\wbis {\rm bf}$. This allows us to study noninterference properties for reversible
systems by using $\wbis{\rm b}$ in a standard process calculus like the one of Section~\ref{sec:proc_lang},
without having to decorate executed actions like in~\cite{PU07,BR23} or store them into stack-based memories
like in~\cite{DK04}.

%
%
\section{Use Case: DBMS Authentication -- Part II}
\label{sec:example_p2}
%
%

The example provided in Section~\ref{sec:example_p1} is useful to illustrate the limitations of weak
bisimilarity when investigating potential covert channels in reversible systems. In particular, it turns out
that $\ms{Auth} \setminus \cala_\calh \not\wbis{\rm b} \ms{Auth} \, / \, \cala_\calh$, i.e., $\ms{Auth}$ is
not BrSNNI, and hence not even BrNDC, SBrSNNI, and SBrNDC by virtue of
Theorem~\ref{thm:branching_taxonomy_1}. As can be seen in Figure~\ref{fig:example_low_views}, the reason is
that, if $\ms{Auth} \, / \, \cala_\calh$ performs the leftmost $\tau$-action and hence moves to state
$r'_3$, from which the only executable action is $l_{\ms{sso}}$, then according to the definition of
branching bisimilarity $\ms{Auth} \setminus \cala_\calh$ can:

	\begin{itemize}

\item either stay idle, but from that state $\ms{Auth} \setminus \cala_\calh$ can then perform actions other
than $l_{\ms{sso}}$ that cannot be matched on the side of $\ms{Auth} \, / \, \cala_\calh$;

\item or perform two $\tau$-actions thereby reaching state $s_3$, but the last traversed state, i.e., $s_2$,
\linebreak is not branching bisimilar to the initial state of $\ms{Auth} \, / \, \cala_\calh$.

	\end{itemize}
\noindent 
In a standard model of execution, where the computation can proceed only forward, the distinguishing power
of branching bisimilarity may be considered too severe, as no practical covert channel actually occurs and
the system can be deemed noninterfering as shown in Section~\ref{sec:example_p1}. Indeed, a low-level user
has no possibility of distinguishing the internal move performed by $\ms{Auth} \, / \, \cala_\calh$ that
leads to $l_{\ms{sso}} \, . \, \ms{Auth}$ from the sequence of internal moves performed by $\ms{Auth}
\setminus \cala_\calh$ that lead to $l_{\ms{sso}} \, . \, \ms{Auth}$ as well. This motivates the fact that,
historically, weak bisimilarity has been preferred in the setting of noninterference.

Now we know that, if we replace the branching bisimulation semantics with the weak back-and-forth
bisimulation semantics, nothing changes about the outcome of noninterference verification. Assuming that the
DBMS allows transactions to be reversed, it is instructive to discuss why BrSNNI is not satisfied by
following the formalization of the weak back-and-forth bisimulation semantics provided in
Section~\ref{sec:bf}. 

After $\ms{Auth} \, / \, \cala_\calh$ performs the run $(r_1, (r_1 \arrow{\tau}{} r'_3
\arrow{l_{\ms{sso}}}{} r_1))$, process $\ms{Auth} \setminus \cala_\calh$ can respond by performing the run
$(s_1, (s_1 \arrow{\tau}{} s_2 \arrow{\tau}{} s_3 \arrow{l_{\ms{sso}}}{} s_1))$. If either process goes back
by undoing $l_{\ms{sso}}$, then the other one can undo $l_{\ms{sso}}$ as well and the states $r'_3$ and
$s_3$ are reached. However, if $\ms{Auth} \setminus \cala_\calh$ goes further back by undoing $s_2
\arrow{\tau}{} s_3$ too, then $\ms{Auth} \, / \, \cala_\calh$ can:

	\begin{itemize}

\item either undo $r_1 \arrow{\tau}{} r'_3$, but in this case $r_1$ enables action $l_{\ms{pwd}}$ while
$s_2$ does not;  

\item or stay idle, but in this case $r'_3$ enables only $l_{\ms{sso}}$, while $s_2$ can go along the path
$s_2 \arrow{\tau}{} s_4 \arrow{l_{\ms{2fa}}}{} s_1$ as well.

	\end{itemize}
\noindent 
This line of reasoning immediately allows us to reveal a potential covert channel under reversible
computing. In fact, let us assume that the transaction modeled by $\ms{Auth}$ is not only executed forward,
but also enables backward computations triggered, e.g., whenever debugging mode is activated. This may
happen in response to some user-level malfunctioning, which may be due, for instance, to the authentication
operation or to the transaction execution. As formally shown above, if the action $l_{\ms{sso}}$ performed
at $r'_3$ after the high-level interaction is undone along with the latter, then the system enables again
the execution of the action $l_{\ms{pwd}}$. This is motivated in our example by the fact that, by virtue of
the transaction rollback, any kind of authentication becomes admissible again. On the other hand, this is
not possible after undoing the action $l_{\ms{sso}}$ performed at state $s_3$, because in such a case the
internal decision of the DBMS of adopting a highly secure mechanism is not reversed. In other words, by
reversing the computation the low-level user can become aware of the fact that the transaction data are
feeding the training set or not.

In the literature, there are several reverse debuggers working in this way like, e.g., UndoDB~\cite{Eng12},
a Linux-based interactive time-travel debugger that can handle multiple threads and their backward
execution. For instance, it is integrated within the DBMS SAP HANA~\cite{UDB} in order to reduce
time-to-resolution of software failures. In our example, by virtue of the observations conducted above, if
the system is executed backward just after performing $l_{\ms{sso}}$, a low-level user can decide whether a
high-level action had occurred before or not, thus revealing a covert channel. Such a covert channel is
completely concealed during the forward execution of the system and is detected only when the system is
executed backward. In general, this may happen when the reverse debugger is activated by virtue of some
unexpected event (e.g., segmentation fault, stack overflow, memory corruption) caused intentionally or not,
and by virtue of which some undesired information flow emerges toward low-level users.

%
%
\section{Conclusions}
\label{sec:concl}
%
%

Our study of branching-bisimilarity-based noninterference properties has established a connection with
reversible computing, in the sense that those properties are directly applicable to reversible systems
expressed in a standard process language with no need of decorating executed actions~\cite{PU07,BR23} or
storing them into stack-based memories~\cite{DK04}. To the best of our knowledge, this is the first attempt
of defining noninterference properties relying on branching bisimilarity so as to reason about information
leakage in reversible systems.

In particular, we have rephrased in the setting of branching bisimilarity the taxonomy \linebreak of
nondeterministic noninterference properties based on weak bisimilarity~\cite{FG01,FR06}. This generates an
extended taxonomy (Figure~\ref{fig:taxon_weak_branching}) that is conservative with respect to the classical
one and emphasizes the strictness of certain property inclusions as well as the incomparability of other
properties. In addition, we have studied preservation (Theorem~\ref{thm:preservation}) and compositionality
(Theorem~\ref{thm:compositionality}) features of the new noninterference properties relying on branching
bisimilarity. Moreover, some ancillary results (Lemmas~\ref{lem:compositionality}
and~\ref{lem:branching_taxonomy}) have been established about parallel composition, restriction, and hiding
under branching bisimilarity, SBrSNNI, and SBrNDC, which elicit general patterns applicable also under weak
bisimilarity, SBSNNI, and SBNDC.

With respect to the earlier version of our study~\cite{EAB23}, the considered process language now supports
recursion -- which has required us to introduce the aforementioned ancillary results and resort to the
notion of branching bisimulation up to $\wbis{\rm b}$ -- and the taxonomy has been extended in such a way to
include a persistent variant of the property of non-deducibility on composition.

We have then shown through a database management system example that potential covert channels arising in
reversible systems cannot be revealed by employing weak bisimulation semantics. Indeed, the higher
discriminating power of branching bisimilarity is necessary to capture information flows emerging when
backward computations are admitted. The correspondence discovered in~\cite{DMV90} between branching
bisimilarity and weak back-and-forth bisimilarity confirms the adequacy of our approach based on the former.

As for future work, we are planning to further extend the noninterference taxonomy so as to include more
expressive properties that take into account also quantitative aspects of process behavior like
in~\cite{ABG04,HMPR21}. To accomplish this for reversible systems, it is necessary to understand whether the
approach of~\cite{DMV90} generalizes to quantitative back-and-forth bisimilarities. Some results for the
probabilistic case can be found in~\cite{EAB24}.

\section*{Acknowledgment}
This research has been supported by the PRIN 2020 project \textit{NiRvAna -- Noninterference and
Reversibility Analysis in Private Blockchains}. We are grateful to Rob van Glabbeek for the valuable
discussions on up-to techniques for branching bisimilarity. We finally thank the anonymous reviewers for
their comments and suggestions.

\bibliographystyle{alphaurl}
\bibliography{biblio}

\end{document}